\documentclass[conference,10pt]{IEEEtran}
\ifCLASSINFOpdf
\else
\fi

\usepackage{textcomp,amssymb}
\usepackage{hyperref}
\usepackage{setspace}
\usepackage{latexsym,fancyhdr,url}
\usepackage{enumerate}
\usepackage[linesnumbered,ruled]{algorithm2e}
\usepackage{algpseudocode}
\usepackage{graphics}
\usepackage{xparse} 
\usepackage{xspace}
\usepackage{multirow}
\usepackage{csvsimple}
\usepackage{balance}
\usepackage{booktabs}
\usepackage{colortbl}
\usepackage{fancyhdr}
\usepackage{makecell}
\usepackage{xcolor, eucal}
\usepackage{array}
\usepackage{subfigure}
\usepackage{graphicx}
\usepackage{caption}
\usepackage{tabularx}
\usepackage{dsfont}
\usepackage{enumitem}
\usepackage{float}
\usepackage{etoolbox}
\usepackage{amsmath}
\usepackage{marvosym}

\usepackage[available,functional,reproduced]{ndssbadges}

\usepackage[most]{tcolorbox}
\usepackage{xcolor}
\newcommand{\grayblock}[1]{\textcolor{gray}{\rule{#1}{1.4ex}}}

\IEEEoverridecommandlockouts

\definecolor{gray0}{gray}{0.9}

\usepackage[most]{tcolorbox}
\newcommand\rev[1]{{\color{black}  #1}}

\def \toolname{$\mathtt{PrivCode}$\xspace}

\begin{document}
%
\title{\toolname: When Code Generation Meets Differential Privacy}



%


\author{\IEEEauthorblockN{Zheng Liu, Chen Gong\textsuperscript{\Letter}, Terry Yue Zhuo\IEEEauthorrefmark{2},
Kecen Li, Weichen Yu\IEEEauthorrefmark{3}, Matt Fredrikson\IEEEauthorrefmark{3},
Tianhao Wang}
\IEEEauthorblockA{University of Virginia}
\IEEEauthorblockA{\IEEEauthorrefmark{2}Monash University and CSIRO's Data61}
\IEEEauthorblockA{\IEEEauthorrefmark{3}Carnegie Mellon University}

\thanks{\textsuperscript{\Letter} Corresponding Author (Chen Gong). Zheng and Kecen work as independent researchers and remote interns at UVA.}}


\IEEEoverridecommandlockouts
\makeatletter\def\@IEEEpubidpullup{6.5\baselineskip}\makeatother
\IEEEpubid{\parbox{\columnwidth}{
    Network and Distributed System Security (NDSS) Symposium 2026 \\
    23-27 February 2026, San Diego, CA, USA \\
    ISBN 979-8-9919276-8-0 \\
    https://dx.doi.org/10.14722/ndss.2026.240936 \\
    www.ndss-symposium.org \\
}
\hspace{\columnsep}\makebox[\columnwidth]{}}

\maketitle

\begin{abstract}
Large language models (LLMs) have presented outstanding performance in code generation and completion. However, fine-tuning these models on private datasets can raise privacy and proprietary concerns, such as the leakage of sensitive personal information. Differentially private (DP) code generation provides theoretical guarantees for protecting sensitive code by generating synthetic datasets that preserve statistical properties while reducing privacy leakage concerns. However, DP code generation faces significant challenges due to the strict syntactic dependencies and the privacy-utility trade-off.

We propose \toolname, the first DP synthesizer specifically designed for code datasets. It incorporates a two-stage framework to improve both privacy and utility. In the first stage, termed ``privacy-sanitizing'', \toolname generates DP-compliant synthetic code by training models using DP-SGD while introducing syntactic information to preserve code structure. The second stage, termed ``utility-boosting,'' fine-tunes a larger pre-trained LLM on the \textit{synthetic privacy-free} code to mitigate the utility loss caused by DP, enhancing the utility of the generated code. Extensive experiments on four LLMs show that \toolname generates higher-utility code across various testing tasks under four benchmarks. The experiments also confirm its ability to protect sensitive data under varying privacy budgets. We provide the replication package at the
code repository.\footnote{\url{https://github.com/Liuzzyg/PrivCode}}
\end{abstract}


\section{Introduction}

The development of large language models (LLMs) has propelled code intelligence into a new era. Advanced open-source code LLMs, such as StarCoder~\cite{starcoder}, CodeLlama~\cite{codellama}, DeepSeekCoder~\cite{guo2024deepseekcoderlargelanguagemodel}, and CodeStral~\cite{codestral}, have presented phenomenal performance and even rival human capabilities in tasks like code generation~\cite{fan2023large,min2023recent}, code completion~\cite{HUSEIN2025103917}, and program-based mathematical reasoning~\cite{tashtoush2022effect}. To better cater to downstream programming scenarios requiring domain-specific expertise, it is common to fine-tune code LLMs on proprietary and sensitive code datasets~\cite{yang2024federated}. 

Previous works show that LLMs can memorize content from the training dataset and output it during inference~\cite{carlini2022quantifying,carlini2021extracting,nasr2023scalable}. For example, Carlini et al.~\cite{carlini2021extracting} show that the GPT-2 language model~\cite{radford2019language} memorizes and outputs the phone number of an individual named `Peter W' with a crafted prompt. For code LLMs, CodexLeaks~\cite{codexleaks} found that Codex~\cite{chen2021evaluating} can reproduce code snippets in verbatim from its training set that contain Personally Identifiable Information (PII). Ziegler et al.~\cite{pearce2021asleepkeyboardassessingsecurity} found that GitHub Copilot~\cite{Copilot} memorizes and reproduces code from its training data, including sensitive information like outdated API keys.

We aim to protect the privacy of code datasets through differentially private (DP) code generation, an approach that generates artificial data, preserving the statistical properties of real data while protecting individual privacy~\cite{li2024privimage,dptab1,yue-etal-2023-synthetic,gong2025dpimagebench}. Specifically, we leverage DP to provide a theoretical guarantee for limiting privacy leakage in the synthesizer's output. We treat each potential sensitive code snippet as a private individual. The instances of code snippets are presented in Figure~\ref{fig:code snippet}. Section~\ref{subsec:dp_code} defines the DP in code generation. To achieve DP, a straightforward approach is to leverage DP-SGD~\cite{dpsgd} to fine-tune a code LLM. However, several challenges remain to directly adopt this method to code generation.


\begin{itemize}[leftmargin=*]
    \item \textit{Utility decrease by DP fine-tuning}: DP inevitably reduces model utility because the noise required to satisfy DP guarantees can negatively impact learning processes. Synthetic data from DP fine-tuned models often has lower utility than the original dataset, reducing effectiveness in downstream tasks. Even when using parameter-efficient methods like LoRA~\cite{lora}, DP fine-tuning of LLMs requires training a larger number of parameters than traditional code synthesizers, such as CODEFUSION~\cite{singh2023codefusion}. For example, a GPT-3 (175B) model with a 144MB LoRA adapter is significantly larger than the 75MB CODEFUSION diffusion model~\cite{singh2023codefusion}. Consequently, fine-tuning LLMs under a given privacy budget requires more Gaussian noise than fine-tuning traditional code synthesizers~\cite{zhu2019gancoder,singh2023codefusion}, resulting in a greater degradation of its utility~\cite{li2024privimage,dockhorn2022differentially}.

    \item \textit{Strong structural dependencies}: Unlike text, code datasets follow strict syntax, semantic rules, and structural dependencies~\cite{ma2024unveiling}. Noise injection for DP can disrupt syntax or key identifiers, leading to uncompileable or non-functional code. DP text synthesis~\cite{yue-etal-2023-synthetic,pretext} overlooks code-specific positional relationships (e.g., paired ``{\tt if-else}'' blocks), and gradient noise further hampers syntax learning, weakening the usability of synthetic codes. As we show in Section~\ref{sub:utility}, directly training DP code synthesizers built on LLMs using DP-SGD~\cite{dpsgd} can lead to utility bottlenecks and limited adaptability to diverse code structures.
\end{itemize}

To resolve the aforementioned dilemma, we propose \toolname, the first DP synthesizer specifically tailored for code generation. \toolname introduces a two-stage framework, which breaks the learning process into “\textit{privacy-sanitizing}” and “\textit{utility-boosting}.” The first stage focuses on learning the code characteristics under privacy constraints. As previous works have shown~\cite{li2024privimage,dockhorn2022differentially}, the noise scale is positively correlated with the size of model parameters under a fixed privacy budget. Therefore, we use a junior LLM (one with a smaller size of parameters) to mitigate the negative impact caused by DP noise. In particular, we train synthesizers on sensitive code datasets using DP-SGD at this stage. Additionally, to strengthen the learning of structural dependencies information, \toolname introduces a Privacy-free Syntax-Aware (PrivSA) module. It extracts structural tokens from code snippets, embedding termed adversarial code syntactic information, a supplementary objective designed to counteract the disruption of code structure caused by DP noise, directly into the fine-tuning process, solving the problem of strong structural dependencies of code generation. 

The second stage “\textit{utility-boosting}” focuses on boosting the synthetic performance.
We know that the post-processing property of DP ensures that operations on DP-compliant outputs do not introduce additional privacy cost~\cite{dpbook}. Therefore, we then fine-tune a premium LLM (a powerful, larger-parameter model) on the synthetic code produced after the privacy-sanitizing stage to focus on utility refinement. We note that not all synthetic code snippets from the junior LLM are used to fine-tune the premium LLM. Because the junior LLM's generative ability is limited under DP, its synthetic codes may be suboptimal, or even include wrong codes. We therefore filter the synthetic code using the following validation and retain only the high-quality portion for fine-tuning the premium LLM. To ensure the functional correctness of synthetic code, we execute the generated code snippets in a controlled environment to filter out those that fail to run, following the execution validation~\cite{chen2021evaluating,Li_2022}.  To ensure the semantic correctness of synthetic code, we summarize each synthetic code snippet into a natural language description. The semantic similarity between this summary and the original prompt is then used to filter out irrelevant code snippets, following the round-trip validation~\cite{allamanis2024unsupervisedevaluationcodellms,bigcode-evaluation-harness}.
The filtered high-quality synthetic dataset is used to fine-tune a premium LLM without any DP constraints, thereby avoiding utility degradation.
This stage is termed the “\textit{utility-boosting},” mitigating the synthetic performance degradation. 


We evaluate the effectiveness of \toolname by leveraging Qwen2.5-Coder-1.5B~\cite{qwen2} as a junior LLM (a smaller-parameter model) in the privacy-sanitizing stage, while Deepseek-Coder-6.7B-Base~\cite{guo2024deepseekcoderlargelanguagemodel}, Qwen2.5-Coder-7B~\cite{qwen2}, CodeGemma-7B~\cite{codegemmateam2024codegemmaopencodemodels}, and CodeQwen1.5-7B~\cite{codeqwen1.5} as the premium LLMs (a powerful larger-parameter model) in the utility-boosting stage. The utility of synthetic code generated by \toolname are evaluated on well-known benchmarks, including HumanEval~\cite{chen2021evaluating}, MBPP~\cite{austin2021programsynthesislargelanguage}, EvalPlus (which includes HumanEval+ and MBPP+)~\cite{liu2023codegeneratedchatgptreally}, and BigCodeBench~\cite{zhuo2024bigcodebenchbenchmarkingcodegeneration}. Compared to the baselines, \toolname shows improvements of up to 9.2\% in code pass rates for instruction following and up to 14.4\% for code completion, across the four benchmarks. Besides, we fine-tune synthesizers on our constructed datasets containing amount of
real-world PIIs and injected canary samples. Comparing the leakage rates of canary samples in code generated under a privacy budget of $\epsilon = 4$ with those from the no-DP baseline, \toolname achieves a 0\% leakage rate, in contrast to the maximum 100\% leakage rate observed in no-DP methods. Ablation studies are performed to emphasize the importance of incorporating PrivSA module into DP fine-tuning and the evolutionary paradigm. In summary, our contributions are three-fold:
\begin{itemize}[leftmargin=*]
    \item We introduce \toolname, the first DP code generation approach using LLMs, which includes two-stage training, ``privacy-sanitizing'' and ``utility-boosting'' stages.
    \item \toolname introduces the privacy-free syntax-aware DP fine-tuning on sensitive datasets, incorporating adversarial code syntactic information to enhance the code generation capability under DP.
    \item Comprehensive evaluations show that \toolname outperforms the baseline across five benchmarks and achieves performance close to methods without any privacy protection mechanisms.  \toolname empirically shows excellent ability to protect sensitive code.
\end{itemize}

\section{Backgrounds}

This section introduces the brief concepts of LLM-based prompted code generation, abstract syntax trees in code, the privacy leakage concerns in code generation, DP in code generation, and the challenges of DP code generation.

\subsection{LLM-based Prompted Code Generation}
\label{sub:codegen}

\begin{figure}[!t]
    \centering
    \includegraphics[width=0.46\textwidth]{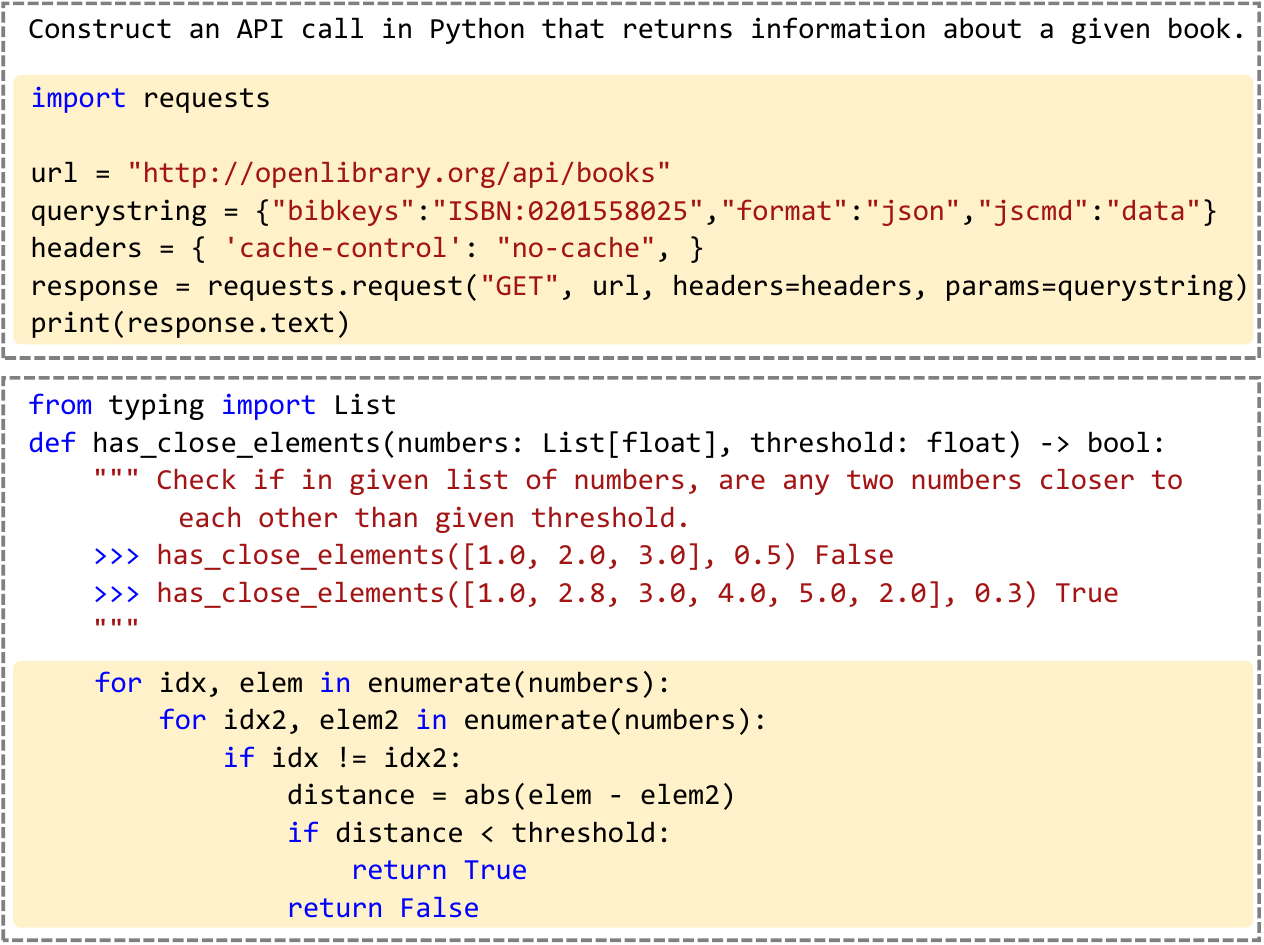}
    \caption{Examples of code snippets. The above part shows \textit{instruction-following data}~\cite{wei2021finetuned}, while the below part shows \textit{code completion data}~\cite{bruch2009learning}. Given the prompt, the highlighted part represents the generated code snippet.}
    \label{fig:code snippet}
    \vspace{-3mm}
\end{figure}

\begin{table*}[h]
\centering
\footnotesize
    \caption{An example of code privacy leakage from the dataset OSS-Instruct PII dataset introduced in Section~\ref{trainset}. The prompt is a public instruction, while the code snippet contains highlighted private information.}
    \vspace{-2mm}
    \label{tab:pii_leakage}
    \begin{tabular}{p{0.06\linewidth} | p{0.05\linewidth} | p{0.81\linewidth}}
   \noalign{\vspace{0.8em}}
    \toprule
         \multicolumn{1}{c}{\textbf{Source}} &  & \multicolumn{1}{l}{\textbf{Text}} \\
    \midrule
       \multirow{12}*{Train Data} & \multirow{4}{*}{Prompt} & Write a Python function that takes a list of dictionaries as input, where each dictionary represents a user with various attributes. The function should return a new list of dictionaries, where each dictionary contains only the 'id', 'firstName', 'lastName', and 'email' attributes of the corresponding user. The function should also add a new attribute 'fullName' to each dictionary, which is the concatenation of 'firstName' and 'lastName'. \\
       \cline{2-3}
       & & \texttt{\detokenize{\`\`\`python\ndef process_users(users):\n    result = []\n    users = [\n    \{\n        "id": 57,\n}} \\ 
       & & \texttt{\detokenize{   ...}} \\
       & Code Snippet & 
       \colorbox{yellow!30}{%
           \begin{minipage}{\linewidth}%
           \texttt{\detokenize{"age": 22,\n        "firstName": "Eli}}\grayblock{1.8em}\texttt{\detokenize{th",\n        "lastName": "Ge}}\grayblock{1.0em}\texttt{\detokenize{ry",\n        "gender": "female",\n        "company": "As}}\grayblock{1.5em}\texttt{\detokenize{ty",\n        "email": "eli}}\grayblock{3.0em}\texttt{\detokenize{try@as}}\grayblock{2.0em}\texttt{\detokenize{ty.com",\n        "phone": "+1 (990) 4}}\grayblock{1.0em}\texttt{\detokenize{-2}}\grayblock{1.0em}\texttt{\detokenize{1",\n        "address": "2}}\grayblock{1.0em}\texttt{\detokenize{ Mi}}\grayblock{2.0em}\texttt{\detokenize{i Place, }}\grayblock{3.0em}\texttt{\detokenize{, New Jersey, 1927",\n}}
           \end{minipage}
           }\\
       & & \texttt{\detokenize{   ...}} \\
       & & \texttt{\detokenize{\n\nprint(process_users(users))\n\`\`\`}} \\
    \bottomrule
    \end{tabular}
    \vspace{-2mm}
\end{table*}

LLMs have revolutionized the field of automated code generation, enabling significant advancements in both proprietary and open-source models. Among proprietary models, Claude-4~\cite{claude4} and GPT-4o~\cite{gpt4o} represent advanced solutions that synthesize accurate, efficient, and contextually appropriate \textit{code snippet}. As shown in Figure~\ref{fig:code snippet}, a \textit{code snippet} is one of the most common elements in code generation tasks~\cite{chen2021evaluating,austin2021programsynthesislargelanguage,zhuo2024bigcodebenchbenchmarkingcodegeneration}. It appears as the code solution in instruction following data or as the segment following a given code prompt in completion data. 


Currently, decoder-only language model architectures, such as Codex~\cite{chen2021evaluating}, have been shown to outperform encoder-decoder models like CodeT5~\cite{wang2021codet5identifierawareunifiedpretrained} on prompted code generation tasks. This class of models follows an autoregressive generation paradigm, predicting the next token based on previously generated tokens. Typically, an autoregressive language model (e.g., GPT-2~\cite{radford2019language}) is trained on an original instruction-code dataset. The model generates code tokens from a prompt using sampling strategies such as Greedy Search~\cite{sutskever2014sequencesequencelearningneural}, Beam Search~\cite{koehn-2004-statistical}, or Top-$k$ Sampling~\cite{fan2018hierarchicalneuralstorygeneration}. Given a public prompt $p$, the probability distribution of the model’s output code sequence $x = (x_1, x_2, \ldots, x_n)$ is:
$$
\mathbb{P}(x \mid p) = \prod_{i=1}^{n} \mathbb{P}(x_i \mid x_1, x_2, \ldots, x_{i-1}, p),
$$
where $\mathbb{P}(x_i \mid x_1, \ldots, x_{i-1}, p)$ represents the probability of generating the $i$-th token $x_i$ given the previously generated tokens $x_{1:i-1}$ and the prompt $p$. 
This paper considers code generation tasks from two common real applications: (1) the instruction-following generation task~\cite{wei2021finetuned} and (2) the code completion task~\cite{bruch2009learning}, termed prompted code generation. In particular, the instruction is a natural language task description in the instruction-following generation task, while the prefix includes the function signature and comment-format task description in the code completion task. The instruction and the prefix code header can both be seen as the prompt $p$, while the output code snippet of each task can be seen as $x$.

\subsection{Abstract Syntax Trees}
Abstract Syntax Tree (AST) is a tree-structured data representation of the syntactic structure of source code snippet~\cite{ast}. It abstracts the syntactic components of a code snippet while omitting details such as parentheses and whitespace that do not affect the syntax, facilitating code analysis and transformation. In an AST, each node represents a structural element of the code snippet, such as expressions, statements, variables, or functions, forming a hierarchical syntactic representation. ASTs are widely utilized in compilers, interpreters, code analysis tools, code transformation, and vulnerability detection. ASTs enable a structured understanding of code, which helps in identifying complex code patterns and semantic relationships. They also support syntax-aware operations, making them ideal for tasks like automated refactoring or source-to-source translation~\cite{baxter1998clone}.

By leveraging the AST, the tokens representing the syntactic structures of a code snippet can be automatically extracted. For example, given a code snippet as follows, 


\begin{lstlisting}[
    language=Python, 
    caption={Example python code snippet.}, 
    numbers=none, 
    xleftmargin=0.6em, 
    linewidth=0.47\textwidth,
    basicstyle=\footnotesize\ttfamily,
    frame=single,
    belowcaptionskip=5pt, 
]
def is_safe_to_move(dest, loc, closeEnemyLocs):   
    moveIn = 1    
    for enemy in rg.locs_around(dest, 
        filter_out=('invalid')):       
        if enemy in closeEnemyLocs:            
            if enemy != loc: moveIn = 0    
    return moveIn == 1
\end{lstlisting}

AST constructs a tree representing the syntactic structure of the code. For instance, a node \textit{FunctionDef} points to structural tokens ``\texttt{\detokenize{def is_safe_to_move(dest, loc, closeEnemyLocs): ...}}'' that define the hierarchical and semantic relationships within the code. This node resides at the top level of the AST and contains child nodes representing the function's components, such as its name, arguments, and body. The parser identifies the function definition by recognizing the ``\texttt{def}'' keyword, followed by the function name and parameter list, and groups the subsequent indented statements into the function body. Each of these syntactic elements corresponds to specific nodes in the AST, allowing precise mapping back to the original source code tokens while capturing their structural and semantic roles.

\subsection{Privacy Leakage in Code Generation}
\label{subsec:privacy_leakage}

Previous studies~\cite{li2024privimage,dptab1,yue-etal-2023-synthetic} have presented that fine-tuning models on sensitive training datasets can lead to the leakage of personal private information. Furthermore, prior works~\cite{ippolito2023preventingverbatimmemorizationlanguage,codexleaks} present that the code generation model also raises the risk of models memorizing and reproducing sensitive code snippet containing sensitive information in the training dataset. SantaCoder~\cite{santacoder} applies PII detection and redaction method only to code snippets within the Stack~\cite{kocetkov2022stack3tbpermissively} code dataset to protect private information. It further categorizes these code PIIs into several common types, including \textit{Emails}, \textit{IP addresses}, \textit{Keys}, \textit{Names}, \textit{Usernames}, and \textit{Passwords}.

Following this, we assume that in prompted code generation, the code snippets that constitute the responses are the primary carriers of private information; rather, the prompts do not pose significant privacy leakage risks, referring to the assumption in previous studies~\cite{mcmahan2017communication,carlini2021extracting}.
We describe the sensitive code snippet and non-sensitive prompt as follows.

\begin{itemize}[leftmargin=*]
    \item Prompts are usually descriptive, general-purpose instructions, or prefix headers to guide the model in generating code and are not tied to particular private contexts, making it suitable for public sharing. Prior work~\cite{DP-RDM} from other tasks of LLMs further supports the validity of this hypothesis.
    \item Code snippets often embed concrete PII, business logic, algorithms, or specific data structures that are part of a company’s intellectual property or private implementation~\cite{romansky2018sourcerer}. Automatically generated code could inadvertently expose private information if it reflects sensitive data from the training corpus (e.g., training data or environmental variables).
\end{itemize}

Hence, this paper focuses on providing DP safeguard for code snippets in code generation contexts while treating the associated prompt as publicly available information.

Table~\ref{tab:pii_leakage} provides real examples of code privacy leakage. The highlighted portions represent private information, such as \textit{age}, \textit{name}, \textit{gender}, \textit{company}, \textit{email}, \textit{phone number}, and \textit{address}. Table~\ref{tab:pii_leakage_examples} presents that the model fine-tuned without any privacy protection directly generates code snippets containing partial private information fragments from the training data. Code snippets generated under DP do not leak private information while still maintaining correctness under privacy budget $\epsilon = \{1,4,10\}$.
We focus on developing a DP code synthesizer to protect private information in code snippets.

\begin{figure*}[t]
    \centering
    \includegraphics[width=0.95\textwidth]{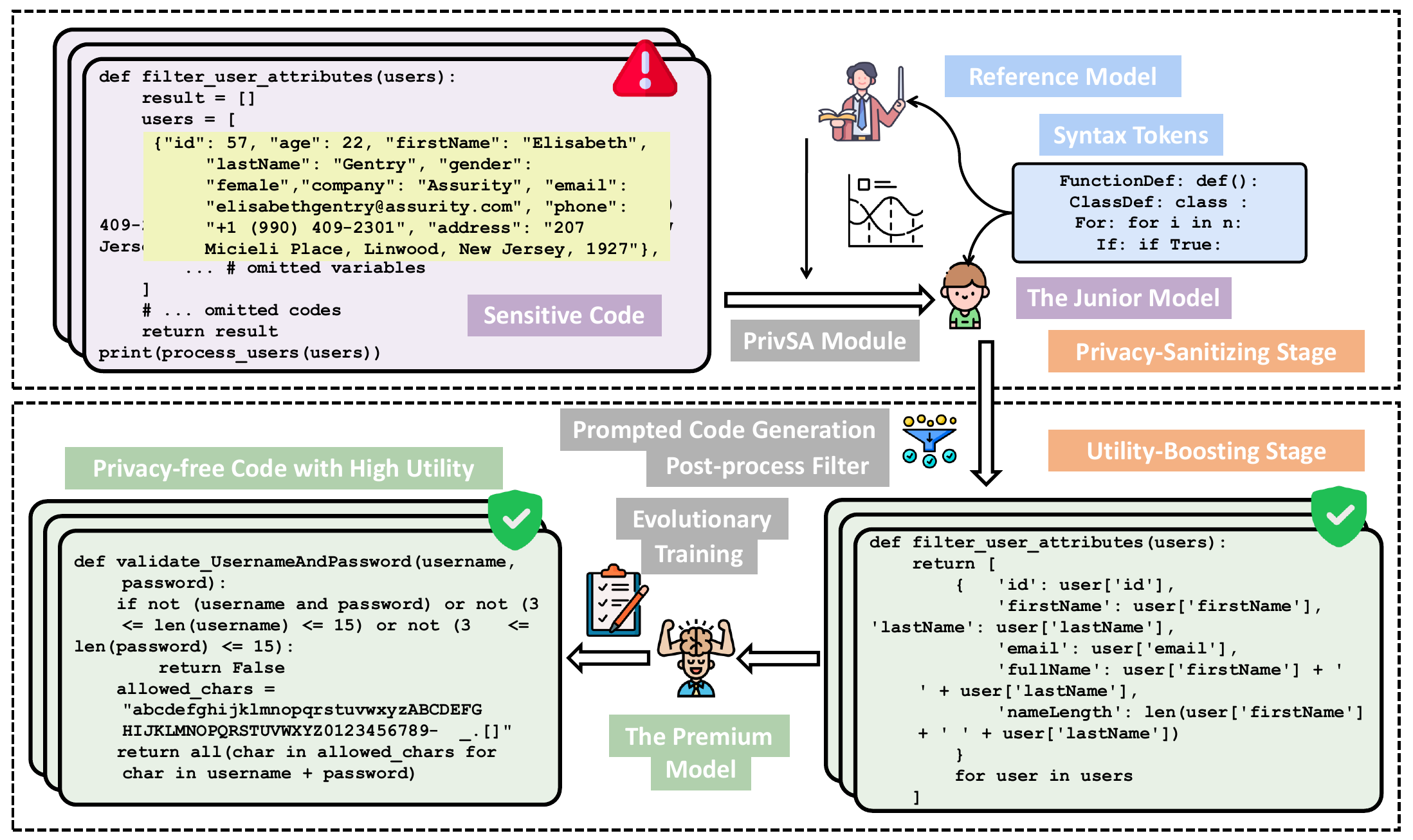} 
    \caption{The workflow of \toolname. The privacy-sanitizing stage fine-tunes the junior LLM with PrivSA module to generate privacy-free code, incorporating semantic information through knowledge distillation and dynamic adjustment. The utility-boosting stage refines the code via post-process filters to fine-tune the premium LLM and produce high-utility code.}
    \label{fig:dpcode}
    \vspace{-4mm}
\end{figure*}

\subsection{Differential Privacy}
\label{subsec:dp_code}
\noindent \textbf{DP Notion.} Differential Privacy (DP)~\cite{dp} is a privacy-preserving framework that limits how much the presence or absence of a single data point in a dataset can affect a model’s output, thus protecting private information. A randomized algorithm $M$ satisfies $(\varepsilon, \delta)$-DP if, for any two neighboring datasets $D$ and $D'$, it holds that:
$$
\Pr[M(D) \in O] \leq e^\varepsilon \Pr[M(D') \in O] + \delta,
$$
where $O$ means any possible output of $M$. The privacy budget $\epsilon$ indicates how much information the algorithm $M$ can reveal; a smaller $\epsilon$ implies stronger privacy. The $\delta$ can be intuitively understood as failure
probability~\cite{dp}. Two datasets $D$ and $D'$ are considered neighbors if one can be obtained from the other by removing or replacing a single data point. This paper is the first one to define the DP in code generation. As explained in Section~\ref{subsec:privacy_leakage}, it focuses on DP code snippets synthesis while treating prompts as public information. We study DP code generation on two code generation tasks: (1) instruction-following code generation and (2) code completion. The notion of neighboring datasets and the protected entity of code snippets differs between these two tasks. We elaborate on the differences as follows.

\begin{itemize}[leftmargin=*]
    \item \textbf{Instruction-following Code Generation:} A data point is defined as a generated code snippet \( x \), derived from a public instruction prompt \( p \). In the DP framework, we define the notion of neighboring datasets $D$ and $D'$ such that they differ by exactly one code snippet \( x \).
    \item \textbf{Code Completion:} A data point is defined as the completed code snippet \( x \), derived from a public prefix code header \( p \). Similarly, by adding or removing one code snippet \( x \) from the datasets \( D \), we obtain neighboring datasets \( D' \).
\end{itemize}

DP provides a theoretical guarantee to measure privacy leakage in the synthesizer’s output, quantifying the risk of revealing private information about real code snippets from synthetic code while still allowing public prompts to facilitate queries and operations.

\vspace{1mm}
\noindent\textbf{\rev{DP Under Code Correlations.}}
\rev{Code contains structural dependencies, such as identifier reuse, logical flow, and imports, that may introduce correlations across tokens. While correlations in text can theoretically weaken DP~\cite{51845}, code snippets are typically much shorter and more modular than long natural-language passages, reducing such effects in practice. Following established practices of applying DP to language-model training~\cite{sinha2025vaultgemmadifferentiallyprivategemma}, we treat each snippet as an indivisible record and rely on DP’s robustness to arbitrary internal structure. Because DP holds under worst-case within-record correlations, snippet-level DP remains a conservative and appropriate privacy formalization for code generation.}

\vspace{1mm}
\noindent\textbf{\rev{DP for Snippet-Level Protection.}}
\rev{Our protected unit is a \emph{code snippet}, which aligns with the typical granularity at which models memorize and potentially leak training content. DP is used not for detecting or redacting PII tokens, but to bound the influence of any snippet on the model’s output, limiting memorization-based extraction attacks~\cite{carlini2021extracting}. The threat model assumes an adversary who queries the trained synthesizer to recover sensitive training snippets. As our goal isn’t to explicitly detect or define PII, we follow prior work~\cite{yue-etal-2023-synthetic}, while our canary tests employ PII-style tokens, these markers are purely diagnostic and do not define the privacy scope. 
The DP applies uniformly to all snippets, including proprietary logic, identifiers, and other sensitive patterns.}


\vspace{1mm}
\noindent \textbf{DP-SGD}. In machine learning, DP-SGD~\cite{dpsgd} incorporates DP into the training of deep learning models, ensuring that the resulting models satisfy formal DP guarantees. Instead of directly using gradients computed from a batch of data samples, DP-SGD first ensures that no single sample overly influences the update step. It achieves this by clipping the $\ell_2$ norm of each per-sample gradient $g_i$ to a fixed threshold $C$:
$g_i \leftarrow \text{clip}(g_i, C) = g_i \left/ \max\left(1, \frac{\|g_i\|_2}{C}\right).\right.$
After clipping, DP-SGD adds noise drawn from a Gaussian distribution $\mathcal{N}(0,\sigma^2 \mathbb{I})$ to the averaged clipped gradients:
$\tilde{g} = \frac{1}{|B|} \left(\sum_{i \in B} g_i + \mathcal{N}(0, \sigma^2 \mathbb{I})\right),$
where $B$ is a batch of samples. By controlling both the clipping norm $C$ and the noise scale $\sigma$, DP-SGD ensures that the influence of each individual data point on the model parameters remains bounded. A privacy accountant tracks how choices of $C$, $\sigma$, and the number of training steps affect the accumulated privacy loss $(\epsilon, \delta)$ throughout training, ensuring that the final model respects the desired privacy budget~\cite{sgm}.

\subsection{Challenges in DP code generation}
\label{subsec:challenges}

This section introduces the challenges in DP code generation from three perspectives. 

\begin{itemize}[leftmargin=*]
    \item \textbf{Strong Dependence on Structure.} Code snippets differ from natural language text, as they are structural data governed by strict syntax and semantic rules with well-defined hierarchies. In DP code generation, noise is injected into the code synthesizers during training, potentially disrupting syntax or key identifiers and rendering code un-compilable or compromising its intended functionality.
    \item \textbf{Privacy–Utility Trade-off. } DP mechanisms inherently lower model utility, as the added noise interferes with learning from sensitive data. This leads to synthetic data from DP fine-tuned models having less utility than the original dataset, reducing effectiveness in downstream tasks.
    \item  \textbf{Evaluation and Verification.} Designing a benchmark tailored to real-world code scenarios, embedding private information that LLMs might memorize and reproduce, is essential to judge how effectively DP mechanisms protect sensitive code snippets. The evaluation and verification methods of DP code generation remain largely unexplored.
\end{itemize}

Code datasets are more structural than text datasets. Injecting noise and applying gradient clipping in DP-SGD disrupt these tokens’ positional and semantic associations, impairing the model’s specialized understanding and generation of code. In particular, we compute entropy 
on 5,000 text samples from the Yelp dataset~\cite{zhang2016characterlevelconvolutionalnetworkstext} and 5,000 code samples from the code part of Magicoder-OSS-Instruct-75K~\cite{magicoder}, obtaining values of 6.738 for text and 1.290 for code. Lower entropy indicates more structured data~\cite{shannon1948mathematical}. 


\section{Methodology}
\label{sec:method}

This section introduces \toolname, the first DP code synthesizer, including adapting existing methods for DP code generation, overview, and technical details.

\subsection{Adapting Existing Methods}
\label{subsec:adapting_exist}

A straightforward approach is to adopt DP text synthesis methods, as text and code share similar properties~\cite{aug-pe,dp-prompt}. Previous DP text generation methods, such as AUG-PE~\cite{aug-pe} and DP-Prompt~\cite{dp-prompt}, progressively guide pre-trained large generative models to produce text synthetic samples resembling sensitive text data. However, these approaches rely solely on pre-trained models without fine-tuning, leading to suboptimal performance in tasks requiring domain-specific expertise. Domain-specific expertise is essential for code generation, as it demands a deep understanding of programming languages, syntax, and functionality. Code outputs must be syntactically correct, semantically meaningful, and executable. 



Another approach to DP text generation
~\cite{dp-opt, yue-etal-2023-synthetic, yu2024privacypreservinginstructionsaligninglarge,carranza2024syntheticquerygenerationprivacypreserving} involves directly training the synthesizer on sensitive text datasets using DP-SGD, making it seemingly adaptable for DP code generation. However, unlike text, code elements such as variables, functions, and control structures are interconnected through strict syntax rules, often reflected in positional relationships (e.g., paired if-else blocks). Traditional methods, without considering such syntax dependencies, make it difficult to learn complex code structures. Besides, the Gaussian noise added to gradients disrupts the learning of fine-grained token-level syntax information and context dependencies, exacerbating errors in the DP synthesizer's understanding of code syntax and semantics. 

\subsection{Overview}
\label{subsec:overall}
This paper proposes \toolname, addressing the challenges discussed in Section~\ref{subsec:challenges}. As the excellent synthesis capabilities of code LLMs across diverse datasets, especially in code generation~\cite{codellama,codeqwen1.5}, \toolname uses code LLMs as the foundational synthesizer. We summarize the high-level contributions of \toolname as follows.

\begin{itemize}[leftmargin=*]
    \item \textbf{Privacy-free Syntax-aware DP fine-tuning.} As elaborated in Section~\ref{subsec:adapting_exist}, code's structural rules are essential in DP code generation. Besides, the noise introduced by DP-SGD further impedes the synthesizer's understanding of the code dataset. \toolname proposes syntax-aware DP fine-tuning, which leverages the syntax information in code datasets to capture the structural correctness intrinsic to code better, thereby enhancing DP code generation. Notably, we extract syntax information from the probability distribution of code tokens, which does not violate DP.  
    
    \item \textbf{Evolutionary paradigm.} Over-relying on one-stage fine-tuning of large-parameter LLMs for DP code generation can lead to excessive utility degradation due to the large Gaussian noise~\cite{li2024privimage,dockhorn2022differentially}. By leveraging LLMs within a two-stage training framework, defined as the evolutionary paradigm, \toolname effectively focuses on utility improvement after meeting DP guarantee.
    In the first stage, we generate synthetic code with DP guarantees. Then, leveraging the post-processing property of DP, we fine-tune a more powerful model on the sanitized outputs without incurring additional privacy cost. 
\end{itemize}

    As shown in Figure~\ref{fig:dpcode}, \toolname adopts a two-stage approach to separate privacy preservation and generation optimization. The first stage, the “privacy-sanitizing” stage, guides the synthesizer in generating synthetic code under DP. A key property of DP is that operations performed on DP-compliant outputs do not incur additional privacy loss~\cite{dpbook}. This allows synthetic code generated under DP constraints to be further refined without reapplying to privacy mechanisms. 
    In the second stage, termed the “utility-boosting” stage, a larger and more capable LLM is available, as we do not need to consider the privacy here. This premium LLM is fine-tuned on the synthetic code generated by the synthesizer in the “privacy-sanitizing” stage. Then, this stage mitigates the performance degradation caused by DP, enhancing generation quality.

\subsection{Technical Details}
\label{sub:tech}

\noindent \textbf{Privacy-Sanitizing Stage.}
This paper introduces the Privacy-free Syntax-Aware (PrivSA) module, which leverages syntax information in code datasets to enhance the DP code synthesizer's ability to learn code structures. Algorithm~\ref{alg:PrivSA_module} presents the processes of PrivSA module. PrivSA module first separately extracts structure tokens from the junior LLM $M_\text{j}$ and the reference pre-training LLM $M_\text{rf}$. A reference LLM $M_\text{rf}$ is a parameter-frozen model with the same architecture and weights as $M_\text{j}$. As presented in Line~\ref{line:extract}, the input is a code snippet sequence, denoted as \textit{S}. We extract the structural tokens of the code snippet in \textit{S} using the AST. The sequence \textit{S} is parsed into an AST, after which structured syntax nodes are filtered out. The type of each node and its corresponding original code tokens are then extracted and returned. This process enables to incorporate structural and syntactic information.



\begin{algorithm}[!t]
\SetKwFunction{ASTTokenExtractor}{ASTTokenExtractor}
\SetKwFunction{ComputeKLLoss}{ComputeKLLoss}
\SetKwFunction{ComputeLambda}{ComputeLambda}
\caption{The workflow of PrivSA module}
\label{alg:PrivSA_module}
\KwIn{Input sequence $S$, junior LLM $M_\text{j}$, reference LLM $M_\text{rf}$, upper bound $\lambda_{\text{max}}$, lower bound $\lambda_{\text{min}}$, decay rate $\alpha$, step interval $\Delta t$ }
\KwOut{Fine-tuned model $M^\text{DP}_\text{j}$}

def \ASTTokenExtractor$(S):$\\
\Indp 
$\mathcal{T} \gets \text{ast.parse}(S)$\;  
$\mathcal{N} \gets \{ n \in \mathcal{T} \mid \text{is\_structural}(n) \}$\;  
return $\{ (\text{type}(n), S[\text{pos}(n)]) \mid n \in \mathcal{N} \}$\;
\Indm

\For{$t \in [T]$}{
    Sample a batch $L_t$ with probability $L/N$\;
    \tcp{AST Tokens Extraction}
    $t_{\text{s}} \gets \ASTTokenExtractor(S)$\label{line:extract}\;
    $p \gets M_\text{j}(t_{\text{s}})$\label{line:computertoken1}\;
    $p^\prime \gets M_\text{rf}(t_{\text{s}})$\label{line:computertoken2}\;
    \tcp{Loss computation}
    $L_{\text{KL}} \gets \ComputeKLLoss(p, p^\prime)$\label{line:klloss}\;
    $t^\prime = \lfloor t / \Delta t \rfloor \cdot \Delta t$\label{line:totalloss1}\\
    $\lambda = \lambda_{\text{min}} + (\lambda_{\text{max}} - \lambda_{\text{min}}) \cdot e^{-\alpha \cdot t^\prime}$\label{line:totalloss2}\\
    \tcp{Gradient Update}
    Update $M_\text{j}$ to minimize Eq. (\ref{eq:total_loss}) by using DP-SGD;
}
\Return $M^\text{DP}_\text{j}$.
\end{algorithm}

As presented in Lines~\ref{line:computertoken1}-\ref{line:computertoken2}, we input the extracted structure tokens $t_{\text{s}}$ into $M_{\text{j}}$ and $M_{\text{rf}}$ to obtain the probability distribution $p$ under the current parameters of $M_{\text{j}}$, as well as the ideal probability distribution $p^\prime$ of $M_{\text{rf}}$. The extent to which $p$ deviates from $p^\prime$ is calculated by the KL divergence, as shown in Line~\ref{line:klloss}, $\text{KL}(P \parallel Q) = \sum_{x \in \mathcal{X}} P(x) \log \frac{P(x)}{Q(x)},$where $P$ represents $p$, and $Q$ represents $p^\prime$. The KL divergence loss $\mathcal{L}_{\text{KL}}$ is scaled by a hyper-parameter $\lambda$ and combined with the cross-entropy loss $\mathcal{L}_{\text{CE}}$ -- the standard loss function for language models. The objective of the privacy-sanitizing stage is,
\begin{equation}
\mathcal{L}_{\text{total}} = \mathcal{L}_{\text{CE}} + \lambda \cdot \mathcal{L}_{\text{KL}}.
\label{eq:total_loss}
\end{equation}
We emphasize that the $\mathcal{L}_{\text{KL}}$ is based solely on the probability distribution over extracted structural tokens. And these probability distributions depend only on the internal relationships among widely used public structural tokens, e.g., the ``\texttt{range}'' token in ``\texttt{for...in range(n)}'', and do not rely on any relationship with specific code snippets. \textit{It does not involve access to the sensitive training dataset and include additional privacy cost}. Moreover, the gradients of Equation~\ref{eq:total_loss}, which includes the KL divergence term, are still protected by gradient clipping and noise addition under DP-SGD. Therefore, the PrivSA module does not introduce any additional privacy loss.

As shown in Section~\ref{sub:ablation}, using a constant $\lambda$ throughout training diminishes the effect of this adversarial information as training iterations increase. The model eventually treats it as a regularization term, leading to results similar to those without incorporating syntactic information. To address this issue, as shown in Line~\ref{line:totalloss1}-\ref{line:totalloss2}, we set $\lambda$ as an exponentially decaying hyper-parameter:
\begin{equation}
\lambda = \lambda_{\text{min}} + (\lambda_{\text{max}} - \lambda_{\text{min}}) \cdot \exp\left(-\alpha \cdot \left\lfloor \frac{t}{\Delta t} \right\rfloor \cdot \Delta t \right),
\end{equation}
where $\lambda_{\text{min}}$ and $\lambda_{\text{max}}$ are the bounds, $\alpha$ controls the decay rate, $t$ is the actual training step, and $\Delta t$ determines the effective step interval. During the early stages of fine-tuning, leveraging the pre-training model's capabilities, the structural token position embeddings extracted earlier are better utilized to produce probability distributions closer to the actual tokens, playing a major role in parameter updates. In the later stages of fine-tuning, $\lambda$ gradually decreases to a predefined lower bound $\lambda_{\text{min}}$, stabilizing the training process. Then, the DP code synthesizer updates its parameters using DP-SGD. 

\vspace{1mm}

\noindent \textbf{Utility-Boosting Stage.} Introducing DP into training often results in a decline in the utility of synthetic datasets~\cite{li2024privimage,dpdm}. To mitigate this, using the post-processing property of DP to refine synthetic data for downstream tasks has become a widely adopted paradigm like previous works~\cite{yue-etal-2023-synthetic,carranza2024syntheticquerygenerationprivacypreserving}. As discussed in Section~\ref{subsec:overall}, \toolname initially fine-tunes a junior LLM with light parameters on sensitive code, as DP significantly impacts the synthesizer's learning process. After privacy-sanitizing training, the junior LLM $M^\text{DP}_\text{j}$ is prompted to generate privacy-free code snippets. Leveraging the post-processing property of DP~\cite{dpbook}, no additional privacy budget is consumed when processing these synthetic code snippets. This dataset is then used to fine-tune a premium LLM $M_\text{p}$ with a larger parameter size, enhancing the utility of the synthetic code. \toolname generates synthetic code using public prompts as conditions, enabling the creation of code snippets tailored to specific languages, functionalities, styles, lengths, and structures based on the prompt's task instructions. This approach results in better correctness and usability code snippets compared to other methods. Furthermore, generating diverse code snippets for the same prompt ensures that the number of generated code snippets meets or exceeds the number of provided public prompts.


However, using the generated code snippets directly for training the premium LLM can result in an unavoidable decline in the performance of the fine-tuned model. This is because some of the snippets are \textit{dirty code snippets}, which may contain incorrect code logic, tokens that do not conform to the programming language's syntax, extra tokens beyond the complete code, or even mixed natural language tokens.

\begin{figure}[!t]
    \centering
    \includegraphics[width=1\columnwidth]{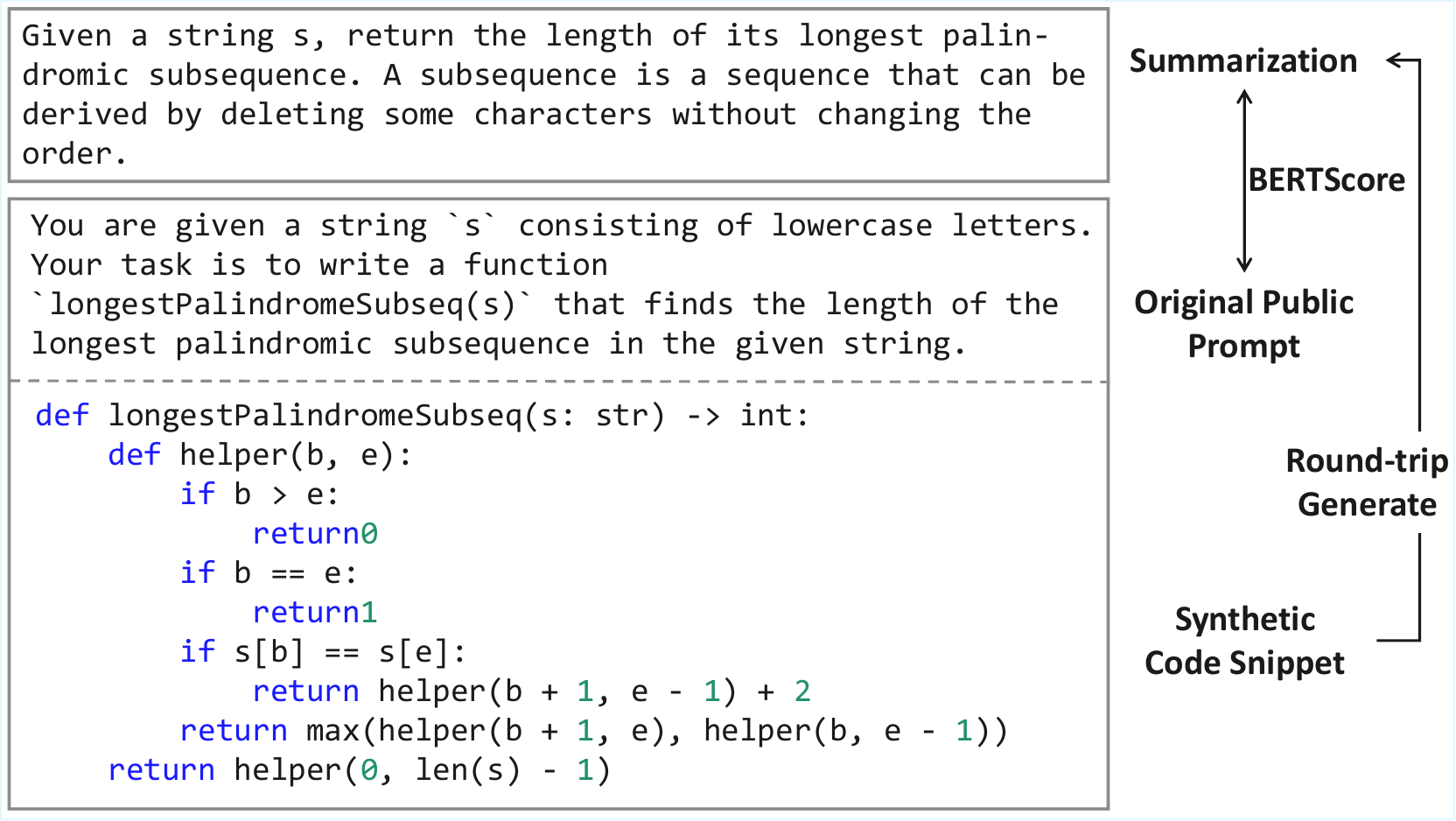}
    \caption{The workflow of round-trip validation. We prompt a round-trip LLM with synthetic code snippets to generate a natural language summarization.
    We use BERTScore as a metric to measure the semantic similarity between the summarization and the original prompt.}
    \label{fig:round-trip}
    \vspace{-3mm}
\end{figure}

\vspace{1mm}
\noindent \textit{Execution Validation.}  Execution validation is a common strategy to ensure that the generated code is syntactically correct and satisfies the target requirements~\cite{chen2021evaluating,Li_2022}. It involves executing the generated code snippets in a controlled environment to verify their correctness. 
\rev{We perform executable validation by combining heuristic language identification, environment capability probing~\cite{guo2011cde}, and sandboxed multi-language runtime execution~\cite{yee2010native}. Based on runtime signals, compiler diagnostics, dependency checks, and heuristic mismatch detection, we classify failure causes into five major categories as follows. 
\begin{itemize}[leftmargin=*]
    \item Environment error refers to issues caused by incorrect or missing dependencies, invalid API usage, unavailable system resources, or non-existent file paths.
    \item Compile error indicates that the code fails to compile or initialize the interpreter.
    \item Runtime error refers to failures occurring after a successful compilation or interpreter startup.
    \item Language mismatch captures cases where the generated code is inconsistent with specified programming languages.
    \item Others include empty code generation, timeout failures, or any unspecified abnormal behavior.
\end{itemize}
Appendix~\ref{sub:execution_validation_appendix} provides explanations and implementation, including statistical breakdowns of practical failure causes.}

As presented in Lines~\ref{algo:ExecutionValidator1}-\ref{algo:ExecutionValidatorend} of Algorithm~\ref{alg:dp_codegen}, for each code snippet $c_\text{i}$ in the input dataset $D_\text{i}$, if the execution fails, the snippet is filtered out. The remaining code snippets $p_\text{e}$ are combined with their corresponding public prompts $p_\text{i}$ as new samples. Only code snippets that pass the execution validation are retained for further processing.

\begin{algorithm}[!t]
\caption{The Workflow of \toolname}
\label{alg:dp_codegen}
\SetKwFunction{ExecutionValidator}{ExecutionValidator}
\SetKwFunction{RoundTripValidator}{RoundTripValidator}
\SetKwFunction{BERTScore}{BERTScore}
\KwIn{Sensitive dataset $\mathcal{D}_\text{s}$ consists of public prompt and sensitive code pairs $\{(\mathcal{P}, 
    \mathcal{C}_\text{s})\}$,
    junior LLM $M_\text{j}$, 
    premium LLM $M_\text{p}$, 
    code executor $E$, 
    round-trip test model $M_\text{r}$, 
    threshold $\tau_{s}$.
}
\tcp{Fine-Tuning with PrivSA }
Fine-tune $M_\text{j}$ on $\mathcal{C}_\text{s}$ using PrivSA module, obtaining the DP code synthesizer $M^\text{DP}_\text{j}$\;\label{algo:dpcode}

\tcp{Prompted Code Generation}\label{algo:dpcode6}
$\mathcal{D} \leftarrow \emptyset$\;\label{algo:dpcode1}
\For{$p \in \mathcal{P}$}{
    Prompt $M^\text{DP}_\text{j}$\ using $p$ to generate privacy-free code snippet $c^\prime$;
    $\mathcal{D}$ = $(p,c^\prime) \cup \mathcal{D}$
}\label{algo:dpcode2}

\tcp{Post-Processing Filter}
\SetKwProg{Fn}{def}{:}{}
\Fn{\ExecutionValidator{$\mathcal{D}_{i}$}\label{algo:ExecutionValidator1}}{
    $\mathcal{D}_{e} \gets \emptyset$\;
    \ForEach{$(p_{i}, c_{i}) \in \mathcal{D}_{i}$}{
        \If{$E(c_{i}) = \text{true}$}{
            $c_{e} \gets c_{i}$, $d_{e} \gets (p_{i}, c_{e})$\;
            $\mathcal{D}_{e} \gets \mathcal{D}_{e} \cup \{d_{e}\}$\;
        }
    }
    \Return $\mathcal{D}_{e}$\label{algo:ExecutionValidatorend}.
}

\Fn{\RoundTripValidator{$\mathcal{D}_{e}$}}{
    $\mathcal{D}_\text{f} \gets \emptyset$\;
    \ForEach{$(p_{i}, c_{e}) \in \mathcal{D}_{e}$}{
        $p_{r} \gets M_{r}(c_{e})$\;\label{algo:RoundTripValidator}
        \If{\BERTScore{$p_{i}, p_{r}$\label{algo:RoundTripValidator1}} $> \tau_{s}$}{
            $\mathcal{D}_{f} \gets \mathcal{D}_{f} \cup \{(p_{i}, c_{e})\}$\;
        }\label{algo:RoundTripValidator2}
    }
    \Return $\mathcal{D}_{f}$\;
}

$\mathcal{D_\text{e}} \leftarrow \ExecutionValidator(\mathcal{D}) $\;\label{algo:dpcode3}
$\mathcal{D}_\text{f} \leftarrow \RoundTripValidator(\mathcal{D}_\text{e})$\;\label{algo:dpcode4}

Fine-tune $M_\text{p}$ on $\mathcal{D}_\text{f}$\ without DP\;\label{algo:dpcode5}

\Return{$M_\text{p}$}
\end{algorithm}

\vspace{1mm}
\noindent \textit{Round-Trip Validation.}
Even after verifying the correctness of the code itself, there remains a potential issue: the generated code may not match the corresponding instruction well, essentially producing irrelevant or inadequate answers. In other words, the code snippet might fail to satisfy the task requirements of the instruction. To handle this, prior works~\cite{allamanis2024unsupervisedevaluationcodellms,muennighoff2024octopack} propose round-trip validation, requiring the model to make predictions (e.g., using natural language to describe some code), provide feedback based on those predictions (e.g., synthesizing code from the predicted description), and check whether this round-trip process leads to code that is semantically equivalent to the original input, thereby eliminating the need for manual inspection.

Directly comparing generated code with the original sensitive code would introduce additional privacy leakage. We adopt a dual approach by employing a powerful pre-training model $M_{r}$ to summarize the generated code into natural language descriptions in Line~\ref{algo:RoundTripValidator}. Figure~\ref{fig:round-trip} presents the workflow of round-trip validation. The semantic similarity between this summarization and the original prompt is computed using BERTScore~\cite{zhang2019bertscore}, which reflects the round-trip correctness of the generated code, i.e., its ability to follow the instructions effectively. In Lines~\ref{algo:RoundTripValidator1}-~\ref{algo:RoundTripValidator2}, we use a carefully chosen hyper-parameter $\tau_{sim}$ as a threshold to filter out samples with low round-trip correctness.

\vspace{1mm}
\noindent \textit{Fine-tuning on Filtered Synthetic Codes.} \toolname fine-tunes the premium LLM $M_\text{p}$ on the filtered synthetic code snippets using the regular SGD. The resulting well-tuned LLM is capable of generating high-utility synthetic code while ensuring privacy protection of sensitive code.

\section{Experimental Setup}
\label{sec:exp_setup}

This section presents the datasets, baselines, and metrics to assess the utility and privacy protection for DP synthesizers.

\subsection{Utilized Datasets}
\label{trainset}
Our experiments involve two training datasets as follows.
We provide more details of studied datasets in Appendix~\ref{sup:dataset_details}.

\noindent{\textbf{Magicoder-OSS-Instruct-75K.}} Magicoder-OSS-Instruct-75K is a dataset created by Magicoder~\cite{magicoder} using its OSS-Instruct method for instruction fine-tuning, includeing a large collection of task-code instruction pairs.

\noindent{\textbf{OSS-Instruct PII Dataset.}} Our constructed instruction-following dataset from the PII dataset~\cite{santacoder}, containing PII-like private tokens in code snippets.



\subsection{Baselines} 
\label{sub:baselines}

To the best of our knowledge, this is the first work to investigate DP code generation. We construct the following three methods as baselines. We provide more details in Appendix~\ref{sup:baselines}.

\begin{itemize}[leftmargin=*]
    \item \textbf{NonDPFT.} This method directly fine-tunes the premium LLM in \toolname with LoRA~\cite{lora}, on sensitive code without DP. Without privacy constraints, it is expected to achieve the best utility compared to other methods in downstream tasks. However, the absence of DP increases the risk of privacy leakage from synthetic code.
    
    \item \textbf{DPFT.} This baseline directly fine-tunes the premium LLM on a sensitive dataset using DP-LoRA~\cite{dplora} under DP. By leveraging DP-SGD~\cite{dpsgd}, DP-LoRA enhance privacy while adding minimal DP noise through fine-tune only parameter-efficient LoRA adapters. However, the added noise inevitably reduces the utility of the synthetic code.
    
    \item \textbf{\rev{DP-Adapter.}} \rev{DP-Adapter~\cite{xu2024dp} applies DP to standard Adapter tuning~\cite{houlsby2019parameter} by freezing the backbone and updating only the small projection modules. Adapter gradients are optimized via DP-SGD to ensure $(\epsilon,\delta)$-DP.}
    
    \item  \textbf{JFT.} This method proposes a two-stage fine-tuning framework, termed Just Fine-tune Twice (JFT)~\cite{jft}. In the first stage, sensitive tokens in the training data are identified using a secret detector and masked, after which the model is fine-tuned on this redacted data without privacy noise. In the second stage, the model is further fine-tuned on the original private data using DP-SGD.    

\end{itemize}

We also list variants of \toolname that we compare in ablation studies as follows:

\begin{itemize}[leftmargin=*]
    \item  \textbf{NonASTPrivCode.} This baseline does not incorporate the code's syntactic knowledge when fine-tuning the junior model $M_\text{j}$ using DP-SGD without PrivSA module.
    \item  \textbf{StableASTPrivCode.} This baseline incorporates the code's syntactic information but sets the KL divergence weight hyper-parameter $\lambda$ as a constant in PrivSA module, maintaining a fixed adversarial loss for syntactic information. This explores the importance of our carefully designed exponentially decaying $\lambda$.
    \item  \textbf{NonEvolPrivCode.} This baseline does not use the evolutionary two-stage training framework; instead, it fine-tunes the premium model $M_\text{p}$ only in the privacy-sanitizing stage. The aim is to show the importance of evolutionary training.
    \item  \textbf{NonPostPrivCode.} This baseline fine-tunes the premium model $M_\text{p}$ directly using the synthetic dataset generated by junior model $M_\text{j}$ fine-tuned in the privacy-sanitizing stage, demonstrating the necessity of execution validation and round-trip validation for low-quality generated code.
\end{itemize}

\subsection{Evaluation Benchmarks and Metrics}
\label{sub:eval}

\noindent{\textbf{Utility Evaluation.}} We select well-known benchmarks such as HumanEval~\cite{chen2021evaluating}, MBPP~\cite{austin2021programsynthesislargelanguage}, EvalPlus (including HumanEval+ and MBPP+)~\cite{liu2023codegeneratedchatgptreally}, and BigCodeBench~\cite{zhuo2024bigcodebenchbenchmarkingcodegeneration}, with their both instruct and complete splits for evaluation. These benchmarks are widely used in evaluating synthetic code~\cite{santacoder, magicoder, starcoder, claude4}. 
\rev{We also select Humaneval-X~\cite{zheng2023codegeex}, a multi-programming-language benchmark, to further evaluate the model’s multilingual code generation capabilities. }
We evaluate our model's effectiveness by reporting pass rates using greedy decoding, which reduces the impact of randomness, and evaluation results are reported as the pass@1 score metric~\cite{chen2021evaluating}, meaning that the generated code must successfully compile and produce the expected output during execution. The pass@1 score reflects the accuracy of code language models based on their initial attempt to generate the correct code. A higher pass@1 score the synthetic code is better. \rev{We also report compile and execution pass rates as utility metrics to support the pass@1 score. The compile pass rate denotes the proportion of generated code snippets that can be successfully parsed and whose target functions can be properly defined, excluding those that fail to compile entirely. The execution pass rate measures the proportion of compiled code snippets that fully pass all tests, calculated relative to the number of successfully compiled samples.} We provide the detailed calculated method and more discussions in Appendix~\ref{supp:utility_eval}.

\begin{table}[!t]
\centering
\caption{The selection of junior and premium LLM models leveraged in \toolname. Models are selected based on a balance of code generative ability and parameter size. `Base' means that LLMs are not instruction-tuned.
}
\resizebox{0.46\textwidth}{!}{
\begin{tabular}{l|lcc}
\toprule
\textbf{Model Type} & \textbf{LLM Model} & \textbf{Year} & \textbf{Type} \\
\hline
Junior Model & Qwen2.5-Coder-1.5B~\cite{qwen2} & 2024 & Base \\
\midrule
\multirow{4}{*}{Premium Model} 
& Deepseek-Coder-6.7B-Base~\cite{guo2024deepseekcoderlargelanguagemodel} & 2024 & Base \\
& Qwen2.5-Coder-7B~\cite{qwen2} & 2024 & Base \\
& CodeGemma-7B~\cite{codegemmateam2024codegemmaopencodemodels} & 2024 & Base \\
& CodeQwen1.5-7B~\cite{codeqwen1.5} & 2024 & Base \\
\bottomrule
\end{tabular}%
}
\label{tab:model_intro}
\vspace{-4mm}
\end{table}


\begin{table*}[h!]
\centering
\scriptsize
\caption{\rev{Pass@1 score of \toolname and baselines under $\epsilon=4$ trained using four LLMs as premium models. The bolded data represents the best score, and the gray shaded area indicates \toolname.}}
\label{tab:utility}
\resizebox{1.0\textwidth}{!}{
\begin{tabular}{l|l|cc|cc|cc|cc|cc|cc}
\toprule
\multirow{3}{*}{\textbf{Model}} & \multicolumn{1}{c|}{\multirow{3}{*}{\textbf{Method}}} & \multicolumn{2}{c|}{\textbf{HumanEval}} & \multicolumn{2}{c|}{\textbf{MBPP}} & \multicolumn{2}{c|}{\textbf{BigCodeBench}} & \multicolumn{2}{c|}{\textbf{HumanEval}} & \multicolumn{2}{c|}{\textbf{MBPP}} & \multicolumn{2}{c}{\textbf{BigCodeBench}} \\
\cline{3-14}
& \multicolumn{1}{c|}{} & HE & HE+ & MBPP & MBPP+ & Full & Hard & HE & HE+ & MBPP & MBPP+ & Full & Hard\\
\cline{3-14}
& \multicolumn{1}{@{}c|}{} & \multicolumn{6}{c|}{\textbf{Instruct}} & \multicolumn{6}{c}{\textbf{Complete}} \\
\hline
\multirow{5}{*}{DS-Coder-6.7B}
& NonDPFT & 60.4& 54.9& 75.1& 61.4& 34.5& 9.5& 53.0& 45.7& 73.8& 60.1& 39.5& 8.1
\\
\cline{2-14}
& DPFT & 54.9& 48.2& 57.7& 48.9& 28.4& 4.1& 45.7& 39.0& 65.8& 54.8& 34.5& 8.1
\\
& \rev{DP-Adapter} & \rev{52.3} & \rev{46.1} & \rev{58.4} & \rev{46.6} & \rev{27.9} & \rev{3.8} & \rev{47.8} & \rev{41.3} & \rev{62.9} & \rev{54.1} & \rev{35.0} & \rev{8.4} \\

& JFT & 53.2& 48.0& 60.8& 52.9& 28.8& 5.3& \textbf{49.2}& 42.7& 66.4& 52.9& 31.5& 8.8\\
\cline{2-14}
& \cellcolor{gray0}\textbf{Ours} & \cellcolor{gray0}\textbf{56.1}& \cellcolor{gray0}\textbf{51.2}& \cellcolor{gray0}\textbf{69.3}& \cellcolor{gray0}\textbf{59.0}& \cellcolor{gray0}\textbf{29.6}& \cellcolor{gray0}\textbf{7.4}& \cellcolor{gray0}47.0& \cellcolor{gray0}\textbf{43.3}& \cellcolor{gray0}\textbf{69.0}& \cellcolor{gray0}\textbf{58.2}& \cellcolor{gray0}\textbf{36.0}& \cellcolor{gray0}\textbf{9.4}
\\ 
\hline
\multirow{5}{*}{Qwen2.5-Coder-7B}
& NonDPFT & 80.5& 72.6& 72.5& 61.9& 39.0& 16.2& 65.2& 56.7& 77.2& 62.7& 46.1& 16.2 \\
\cline{2-14}
& DPFT & 59.8& 53.7& 58.7& 51.6& 22.1& 5.4& 36.0& 29.9& 56.3& 46.8& 26.7& 6.1
\\
& \rev{DP-Adapter} & \rev{61.0} & \rev{55.5} & \rev{63.9} & \rev{54.5} & \rev{19.6} & \rev{3.9} & \rev{36.0} & \rev{28.2} & \rev{61.4} & \rev{49.5} & \rev{25.2} & \rev{5.8} \\

& JFT & 63.1& 56.3& 69.2& 60.0& 21.9& 7.9& 38.8& 33.0& 67.6& 54.1& 27.2& 8.0\\
\cline{2-14}
& \cellcolor{gray0}\textbf{Ours} & \cellcolor{gray0}\textbf{66.5}& \cellcolor{gray0}\textbf{61.0}& \cellcolor{gray0}\textbf{78.3}& \cellcolor{gray0}\textbf{64.8}& \cellcolor{gray0}\textbf{22.9}& 
\cellcolor{gray0}\textbf{9.5}& \cellcolor{gray0}\textbf{43.9}& \cellcolor{gray0}\textbf{38.7}& \cellcolor{gray0}\textbf{77.9}& \cellcolor{gray0}\textbf{65.6}& \cellcolor{gray0}\textbf{27.9}& \cellcolor{gray0}\textbf{8.8}
\\ 
\hline
\multirow{5}{*}{CodeGemma-7B}
& NonDPFT & 54.3& 48.2& 64.0& 54.8& 27.4& 8.1& 43.3& 37.2& 64.8& 50.3& 32.4& 8.1
\\
\cline{2-14}
& DPFT & 34.1& 30.5& 56.1& 42.9& 22.6& 5.1& 30.5& 25.6& 45.0& 36.2& 22.7& 7.4
\\
& \rev{DP-Adapter} & \rev{36.2} & \rev{29.3} & \rev{58.4} & \rev{44.5} & \rev{20.3} & \rev{4.8} & \rev{34.7} & \rev{25.6} & \rev{48.8} & \rev{40.1} & \rev{24.3} & \rev{7.7} \\

& JFT & 38.5& 30.8& 61.7& 43.6& \textbf{24.9}& 4.8& 33.2& 26.9& 51.7& 46.6& 24.8& 7.2\\
\cline{2-14}
& \cellcolor{gray0}\textbf{Ours} & \cellcolor{gray0}\textbf{42.1}& \cellcolor{gray0}\textbf{36.6}& \cellcolor{gray0}\textbf{65.6}& \cellcolor{gray0}\textbf{53.7}& \cellcolor{gray0}22.9& \cellcolor{gray0}\textbf{5.4}& \cellcolor{gray0}\textbf{40.2}& \cellcolor{gray0}\textbf{31.7}& \cellcolor{gray0}\textbf{66.1}& \cellcolor{gray0}\textbf{53.7}& \cellcolor{gray0}\textbf{30.0}& 
\cellcolor{gray0}\textbf{8.7}
\\ 
\hline
\multirow{5}{*}{CodeQwen1.5-7B}
& NonDPFT & 64.0& 56.7& 73.5& 61.6& 33.7& 10.8& 59.8& 53.7& 74.3& 60.6& 39.3& 12.1
\\
\cline{2-14}
& DPFT & 44.5& 38.4& 65.8& 55.8& 27.2& 8.8& 43.9& 39.0& 66.4& 54.5& 28.8& 7.4
\\
& \rev{DP-Adapter} & \rev{47.1} & \rev{40.2} & \rev{63.7} & \rev{54.9} & \rev{27.6} & \rev{9.5} & \rev{45.2} & \rev{39.8} & \rev{69.6} & \rev{56.1} & \rev{29.8} & \rev{6.7} \\

 & JFT & 50.6& 42.6& 65.1& 55.0& 26.9& 9.2& 45.8& 41.1& 68.1& 56.9& 32.4& 7.9\\
\cline{2-14}
& \cellcolor{gray0}\textbf{Ours} & \cellcolor{gray0}\textbf{52.4}& \cellcolor{gray0}\textbf{44.5}& \cellcolor{gray0}\textbf{70.6}& \cellcolor{gray0}\textbf{60.1}& \cellcolor{gray0}\textbf{29.1}& 
\cellcolor{gray0}\textbf{10.8}& \cellcolor{gray0}\textbf{48.8}& \cellcolor{gray0}\textbf{41.5}& \cellcolor{gray0}\textbf{72.5}& \cellcolor{gray0}\textbf{61.4}& \cellcolor{gray0}\textbf{35.5}& 
\cellcolor{gray0}\textbf{8.8}
\\
\bottomrule
\end{tabular}
}
\vspace{-2mm}
\end{table*}

\noindent{\textbf{Private Information Protection Evaluation.}} We conduct canary experiments to verify whether sensitive training data are memorized and leaked in the synthetic code snippets generated by \toolname, following the approach of previous work~\cite{yue-etal-2023-synthetic,carlini2022quantifying,kandpal2022deduplicating}. We construct five categories of sensitive canary samples and inject them into the OSS-Instruct PII dataset as a training dataset with various repetition rates, each of which contains a distinct type of PII that represents highly private information~\cite{party2014opinion}. The repetition rate refers to the number of times each canary sample is injected into the training dataset. These canary examples, formed as instruction-following pairs, consist of public prompts containing no private information and corresponding code snippet solutions that include one distinct category of PII. Then, using prompts from the testing set of OSS-Instruct PII dataset, we prompt models to generate code snippets. The number of times that the PII canary sample of each category appears in generated code snippets is used to calculate the leakage rate, \rev{ and category-wise leakage count under different repetition rates. The leakage rate means the distribution of leakage canary categories, while the category-wise leakage count means the number of leakage canary instances of each PII category, both reflecting the ability of methods to protect private information of sensitive datasets.}

\subsection{Implementation}
\label{sup:implementation}

Table~\ref{tab:model_intro} shows that \toolname uses Qwen2.5-Coder-1.5B~\cite{qwen2} as the junior model $M_\text{j}$ in the privacy-sanitizing stage. We select a smaller parameter model as $M_\text{j}$ is to demonstrate that \toolname’s effectiveness is not dependent on the utility of pre-trained $M_\text{j}$ itself. For the round-trip model $M_{rt}$, we use the powerful Llama-3.1-70B-Instruct~\cite{grattafiori2024llama3herdmodels}. We utilize four commonly used code LLMs as premium models $M_\text{p}$: Deepseek-Coder-6.7B-Base~\cite{guo2024deepseekcoderlargelanguagemodel}, Qwen2.5-Coder-7B~\cite{qwen2}, CodeGemma-7B~\cite{codegemmateam2024codegemmaopencodemodels}, and CodeQwen1.5-7B~\cite{codeqwen1.5}. To accelerate training, in all stages of fine-tuning, we utilize DeepSpeed~\cite{deepspeed} combined with LoRA~\cite{lora}, a parameter-efficient fine-tuning technique. 


\section{EMPIRICAL EVALUATIONS}
\label{results}
\vspace{-1.5mm}
This section first compares the effectiveness of \toolname with baseline methods in downstream tasks. Next, we analyze how \toolname protects the sensitive training dataset against privacy leakage. Following that, we examine how hyper-parameters and privacy budgets influence the performance of \toolname. Finally, we conduct ablation studies to highlight the contributions of the PrivSA module, code post-processing filters, and evolutionary training.

\begin{table}[!t]
\centering
\caption{\rev{Compile pass rate (Comp.) and execution pass rate (Exec.) of \toolname and baselines on Qwen2.5-Coder-7B  under $\epsilon=4$ and HumanEval and HumanEval+ benchmarks of instruction-following.}}
\vspace{-1mm}
\label{tab:utility_compile_execution_pass_rate}
\setlength{\tabcolsep}{3.2mm}{
\resizebox{0.49\textwidth}{!}{
\begin{tabular}{l|ccc|ccc}
\toprule
\multirow{2}{*}{\textbf{Method}} 
& \multicolumn{3}{c|}{\textbf{HumanEval}} 
& \multicolumn{3}{c}{\textbf{HumanEval+}} \\
\cline{2-7}
& Comp. & Exec. & Pass@1 
& Comp. & Exec. & Pass@1 \\
\midrule
NonDPFT   & 100.0   & 80.5 & 80.5 & 100.0 & 72.6 & 72.6 \\
\hline
DPFT      & 95.1  & 62.8  & 59.8 & 95.1  & 56.4   & 53.7 \\
DP-Adapter& 97.0    & 61.6 & 61.0   & 95.7  & 58.0   & 55.5 \\
JFT       & 98.8  & 64.2& 63.1 & 97.0  & 57.9   & 56.3 \\
\hline
PrivCode\cellcolor{gray0} & 100.0  \cellcolor{gray0} & 66.5 \cellcolor{gray0} & 66.5 \cellcolor{gray0} & 100.0 \cellcolor{gray0} & 61.0 \cellcolor{gray0}  & 61.0 \cellcolor{gray0} \\
\bottomrule
\end{tabular}
}}
\vspace{-5mm}
\end{table}

\subsection{The Utility of Synthetic Codes}
\label{sub:utility}

\noindent \textbf{Experiment Design.} This experiment selects Magicoder-OSS-Instruct-75K dataset as training dataset. We randomly sampled 20,000 examples to train the junior LLM $M_\text{j}$ and 55,000 examples for prompted code generation using $M_\text{j}$. After post-processing and validation, the evolutionary training dataset is generated to train the premium LLM $M_\text{p}$. In PrivSA module, we set the upper bound of the KL loss weight hyper-parameter $\lambda_{\text{max}}$ to 1000, the lower bound $\lambda_{\text{min}}$ to 0.01, the decay rate $\alpha$ to 0.01, and the step interval $\Delta t$ to 20. We use slack round-trip parameters, generating 20 round-trip samples per code and setting the similarity threshold $\tau_{sim}$ to 0.88 to ensure the selection of high-quality synthetic datasets. 


\begin{figure*}[t]
    \centering
    \includegraphics[width=0.98\textwidth]{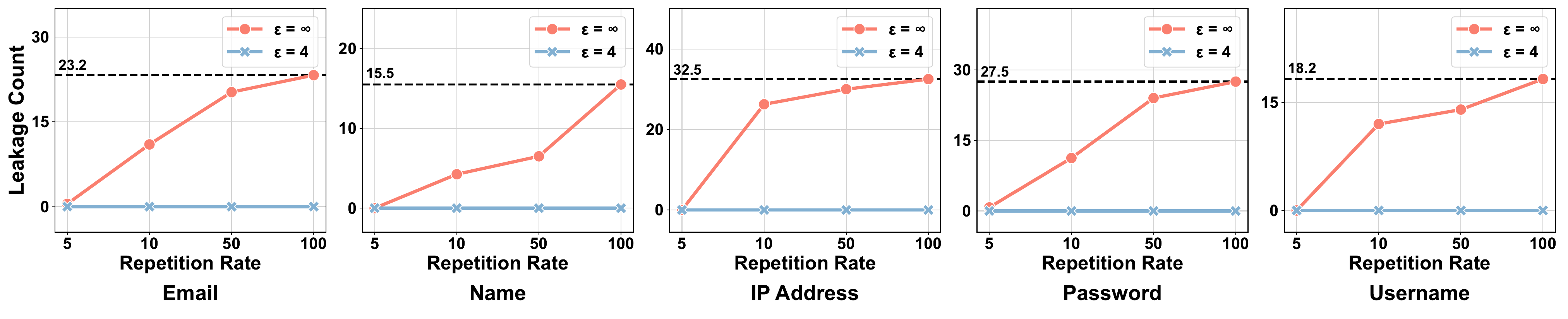} 
    \vspace{-1em}
    \caption{\rev{The average category-wise leakage count of four LLMs trained using \toolname under $\epsilon = \{4,\infty\}$. The $\epsilon = \infty$ means training \toolname without DP. Repetition rate is the number of times each canary sample is injected into training datasets.}}
    \label{fig:leakage_count}
    \vspace{-3mm}
\end{figure*}

\noindent \textbf{Result Analysis.} Table~\ref{tab:utility} presents the pass@1 scores of four models fine-tuned with \toolname and baselines under $\epsilon = 4$. We observe that across most benchmarks, PrivCode consistently achieves the highest pass@1 scores compared to DPFT, \rev{DP-Adapter} and JFT, and is close to NonDPFT. \rev{DPFT and DP-Adapter exhibit similar performance, as both apply parameter-efficient fine-tuning directly leveraging DP-SGD. JFT shows moderate improvement due to more carefully designed training techniques.} In the instruct split, among the four models evaluated, \toolname outperforms the best baseline with improvements of \{3.6, 5.8, 9.1, 9.2, 1.5, 2.1\} across the HumanEval, HumanEval+, MBPP, MBPP+, BigCodeBench (full and hard splits) benchmarks. In the complete split, \toolname achieves improvements of \{5.5, 5.7, 14.4, 11.5, 5.2, 1.0\} over the best baseline on the same set of benchmarks. These results indicate that \toolname mitigates the adverse impact of DP mechanisms on model utility.

\begin{table}[!t]
\centering
\caption{\rev{Pass@1 score of \toolname and baselines under $\epsilon=4$ across Java, C++, and Rust code generation tasks. ‘Pretrain’ refers to generating code directly using LLMs without any fine-tuning on sensitive code.}}
\label{tab:humaneval-x}
\footnotesize
\resizebox{0.5\textwidth}{!}{
\begin{tabular}{l|ccc|ccc|ccc}
\toprule
\multirow{2}{*}{\textbf{Method}}  &
\multicolumn{3}{c|}{\textbf{Qwen2.5-Coder-7B}} &
\multicolumn{3}{c|}{\textbf{CodeGemma-7B}} &
\multicolumn{3}{c}{\textbf{CodeQwen1.5-7B}} \\
\cline{2-10}
& Java & C++ & Rust & Java & C++ & Rust& Java & C++ & Rust \\
\midrule
Pretrain 
& 44.5& 14.0& 47.0& 29.9& 0.6& 28.0& 35.4& 0.0& 40.2\\
NonDPFT 
& 55.5 & 17.1 & 50.6 
& 42.1 & 17.1 & 31.7
& 54.3 & 22.6 & 39.0 \\
\hline
DPFT   
& 25.0 & 15.9 & 22.0 
& 28.0 & 9.1  & 23.8
& 32.3 & 7.9  & 36.0 \\
JFT   
& 43.9 & 10.4 & 37.7 
& 22.6 & 11.6 & 12.8
& 25.0 & 14.0 & 32.3 \\
\hline
\cellcolor{gray0}PrivCode 
& \cellcolor{gray0}\textbf{57.3} 
& \cellcolor{gray0}\textbf{22.0} 
& \cellcolor{gray0}\textbf{44.5} 
& \cellcolor{gray0}\textbf{42.7}
& \cellcolor{gray0}\textbf{12.8}
& \cellcolor{gray0}\textbf{28.7}
& \cellcolor{gray0}\textbf{54.9}
& \cellcolor{gray0}\textbf{17.1}
& \cellcolor{gray0}\textbf{42.7} \\
\bottomrule
\end{tabular}
}
\vspace{-2mm}
\end{table}

\rev{Table~\ref{tab:utility_compile_execution_pass_rate} shows that, across the HumanEval and HumanEval+ benchmarks of instruction-following tasks, \toolname consistently achieves the highest compile pass rate of 100\%, matching the NonDPFT and surpassing all other baselines. Similarly, \toolname attains the highest execution pass rate compared to baselines, which directly contributes to its superior pass@1 scores. Notably, while other baselines experience a decline in compile pass rate on the more challenging HumanEval+ benchmark—reflecting increased syntax errors, \toolname maintains zero syntax errors in generated code. This validates the crucial effectiveness of \toolname in maintaining code syntax correctness under DP.}

\rev{We also expand experiments to non-Python multi-language code generation tasks, including Java, C++, and Rust code generation tasks as well. Table~\ref{tab:humaneval-x} further shows that \toolname consistently achieves the highest multilingual code generation performance across Java, C++, and Rust. Despite the increased syntactic complexity and stricter compiler constraints in these languages, \toolname outperforms all privacy-preserving baselines across all settings, achieving maximum improvements of $\{22.6, 6.1, 6.7\}$ pass@1 scores on Java, C++, and Rust respectively, and often approaches the NonDPFT performance. These results confirm that \toolname effectively preserves the utility of fine-tuning knowledge.}

We observe that \toolname consistently outperforms the DPFT baseline, and JFT also achieves better performance than DPFT on most benchmarks. We analyze that both \toolname and JFT contain a so-called ``utility-boosting stage.'' For \toolname, this refers to fine-tuning on the privacy-free synthetic dataset without DP protection, whereas for JFT, it involves fine-tuning on the masked sensitive dataset, also without DP constraints. These findings highlight the importance of the two-stage paradigm in enhancing model utility under privacy-preserving settings.

The evolutionary training framework presents some intriguing results. For example, Qwen2.5-Coder-7B and CodeQwen1.5-7B achieve pass@1 scores on some benchmarks that even exceed or match NonDPFT (e.g., 65.6 vs. 62.7, 78.3 vs. 72.5, 64.8 vs. 61.9 on MBPP, and 10.8 vs. 10.8 on BigCodeBench). This is because Qwen2.5-Coder-1.5B is used as the junior model $M_\text{j}$, and the premium models $M_\text{p}$, Qwen2.5-Coder-7B and CodeQwen1.5-7B, share similar model architectures and token embeddings in Qwen series. Previous study~\cite{yang2024smalltolarge} shows that examples exhibiting similar loss trajectories on small models tend to have similar gradient behavior on larger models, making the synthetic dataset produced by $M_\text{j}$ more effective as a high-quality training dataset for fine-tuning $M_\text{p}$ to achieve higher utility.





\begin{table}[!t]
\centering
\caption{The leakage rate of four LLMs trained using \toolname under $\epsilon = \{4,\infty\}$. Repetition rate refers to the number of times each canary sample is injected into the training dataset.}
\resizebox{\columnwidth}{!}{%
\begin{tabular}{l|cc|cc|cc}
\toprule
\multirow{3}{*}{\textbf{Model}} & \multicolumn{6}{c}{Repetition Rate} \\ 
\cline{2-7}
& \multicolumn{2}{c}{5} & \multicolumn{2}{c}{10} & \multicolumn{2}{c}{100} \\
\cline{2-7}
& $\epsilon = \infty$ & $\epsilon = 4$ & $\epsilon = \infty$ & $\epsilon = 4$ & $\epsilon = \infty$ & $\epsilon = 4$    \\ 
\hline
DS-Coder-6.7B & 0\%& 0\%& 40\%& 0\%& 60\%&
0\%\\
Qwen2.5-Coder-7B & 20\%& 0\%& 80\%& 0\%& 80\%&
0\%\\ 
CodeGemma-7B & 0\%& 0\%& 40\%& 0\%& 60\%&
0\%\\ 
CodeQwen1.5-7B & 40\%& 0\%& 100\%& 0\%& 100\%&
0\%\\ 
\hline
\cellcolor{gray!20}\textbf{Average} & \cellcolor{gray!20}15\%&  \cellcolor{gray!20}0\%& \cellcolor{gray!20}65\% & \cellcolor{gray!20}0\%& \cellcolor{gray!20}75\% & \cellcolor{gray!20}0\%\\ 
\bottomrule
\end{tabular}%
}
\label{tab:pii}
\vspace{-5mm}
\end{table}

\subsection{Private Information Protection}
\label{sub:pii}

\begin{table*}[h!]
\centering
\scriptsize
\caption{Average Pass@1 of four LLMs trained using \toolname, evaluated on instruct and complete models of HumanEval, MBPP, and EvalPlus benchmarks. The pass@1 score varies with different max lambda. The bolded data is the best score.}
\vspace{-1mm}
\resizebox{1.0\textwidth}{!}{
\setlength{\tabcolsep}{4.5mm}{
\begin{tabular}{l|l|cc|cc|cc|cc}
\toprule
\multirow{3}{*}{\textbf{Hyper-Parameter}} & \multirow{3}{*}{\textbf{Model}} & \multicolumn{2}{c|}{\textbf{HumanEval}} & \multicolumn{2}{c|}{\textbf{MBPP}} & \multicolumn{2}{c|}{\textbf{HumanEval}} & \multicolumn{2}{c}{\textbf{MBPP}} \\
\cline{3-10}
& \multicolumn{1}{c|}{} & HE & HE+ & MBPP & MBPP+ & HE & HE+ & MBPP & MBPP+ \\
\cline{3-10}
& \multicolumn{1}{c|}{} & \multicolumn{4}{c|}{\textbf{Instruct}} & \multicolumn{4}{c}{\textbf{Complete}} \\
\midrule
\multirow{5}{*}{\textbf{$\lambda_{\text{max}}=1$}} 
& DS-Coder-6.7B & 42.7& 34.1
& 69.6& 58.2& 42.6& 37.8& 67.7& 56.6
\\
& Qwen2.5-Coder-7B & 68.3& 64.0& 63.5& 54.0& 46.3& 40.8& 64.0& 54.2
\\
& CodeGemma-7B & 43.9& 39.6
& 64.8& 53.9& 40.2& 32.3& 62.9& 51.5
\\
& CodeQwen1.5-7B & 43.8& 40.2
& 67.7& 58.2& 49.3& 42.6& 68.7& 59.7
\\
\cline{2-10}
& \cellcolor{gray0}\textbf{Average} & \cellcolor{gray0}49.7& \cellcolor{gray0}44.5
& \cellcolor{gray0}66.4& \cellcolor{gray0}56.1& \cellcolor{gray0}44.6& \cellcolor{gray0}38.4& \cellcolor{gray0}\textbf{65.8}& \cellcolor{gray0}\textbf{55.5}
\\
\midrule
\multirow{5}{*}{\textbf{$\lambda_{\text{max}}=1 \times 10^3$}}& DS-Coder-6.7B & 54.3& 46.3
& 69.9& 59.8& 48.2& 40.2& 68.0& 56.9
\\
& Qwen2.5-Coder-7B & 67.1& 60.4
& 73.0& 61.9& 53.7& 48.2& 62.2& 49.7
\\
& CodeGemma-7B & 39.6& 34.8
& 68.0& 56.9& 42.1& 33.5& 52.1& 44.4
\\
& CodeQwen1.5-7B & 50.6& 45.7
& 69.6& 58.5& 52.4& 45.1& 68.2& 56.6
\\
\cline{2-10}
& \cellcolor{gray0}\textbf{Average} & \cellcolor{gray0}\textbf{52.9}& \cellcolor{gray0}\textbf{46.8}
& \cellcolor{gray0}\textbf{70.1}& \cellcolor{gray0}\textbf{59.3}& \cellcolor{gray0}\textbf{49.1}& \cellcolor{gray0}\textbf{41.8}& \cellcolor{gray0}62.6& \cellcolor{gray0}51.9
\\
\midrule
\multirow{5}{*}{\textbf{$\lambda_{\text{max}}=1 \times 10^5$}}& DS-Coder-6.7B & 35.4& 33.5
& 69.3& 58.2& 46.3& 39.6& 69.5& 57.6
\\
& Qwen2.5-Coder-7B & 73.2& 64.6
& 76.2& 63.8& 46.9& 42.1& 73.8& 63.2
\\
& CodeGemma-7B & 42.1& 37.8
& 62.2& 48.7& 35.4& 27.4& 57.4& 46.3
\\
& CodeQwen1.5-7B & 18.9& 17.7
& 16.1& 14.0& 28.0& 25.0& 36.5& 30.1
\\
\cline{2-10}
& \cellcolor{gray0}\textbf{Average} & \cellcolor{gray0}42.4& \cellcolor{gray0}38.4& \cellcolor{gray0}56.0& \cellcolor{gray0}46.2& \cellcolor{gray0}39.2& \cellcolor{gray0}33.5& \cellcolor{gray0}59.3& \cellcolor{gray0}49.3\\

\bottomrule
\end{tabular}
}}
\label{tab:hyper_lambda}
\vspace{-4mm}
\end{table*}

\noindent \textbf{Experiment Design.} The canary experiment is a widely recognized method that inserts unique, private-like sequences, called canaries, into the training dataset and evaluates whether the model can memorize and reproduce them, assessing the risk of sensitive training information leakage~\cite{kandpal2022deduplicating, carlini2022quantifying,yue-etal-2023-synthetic}.


This experiment uses the OSS-Instruct PII dataset, injected canary samples with varying repetition rates.
The evaluation benchmark and metric are described in Section~\ref{sub:eval}. We feed Llama-3.1-70B-Instruct~\cite{grattafiori2024llama3herdmodels} with ten instruction-following samples from the OSS-Instruct PII dataset covering different task types as few-shot context to generate five types of canary instruction-following samples. In these canary samples, the instructions are privacy-free task descriptions excluding any PII, while the corresponding code snippets contain the distinct type of PII. The five types of PIIs included are: Email, Name, IP Address, Password, and Username, and each canary sample is restricted to containing only one specific type of PII (e.g., \texttt{\detokenize{\`\`\`python\n contact = }}\colorbox{yellow!30}{\texttt{\detokenize{\"sarah.}}\grayblock{5.0em}\texttt{\detokenize{90@gmail.com\"}}}\texttt{\detokenize{\nprint(validate_contact(contact))\`\`\`}}). We present complete five canary samples and more details in Appendix~\ref{sub:canary_samples}. 

\begin{figure}[!t]
    \centering
    \vspace{-1mm}
    \includegraphics[width=1\columnwidth]{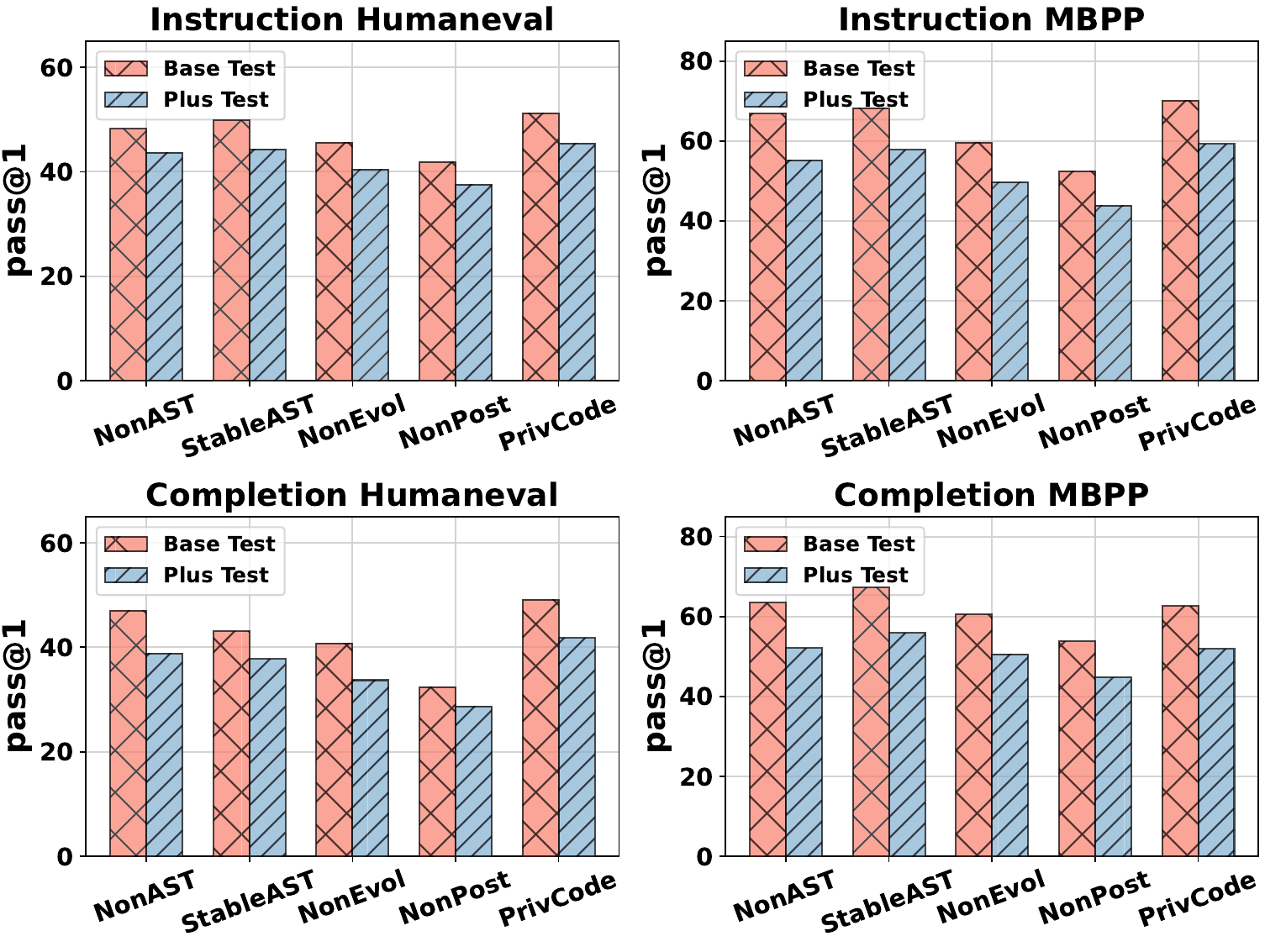}
    \vspace{-5mm}
    \caption{Average pass@1 scores of four models fine-tuned with \toolname and its four variants under $\epsilon=4$. ``Base Test'' refers to HumanEval or MBPP, while ``Plus Test'' refers to HumanEval+ and MBPP+.}
    
    \vspace{-5mm}
    \label{fig:ablation}
\end{figure}



The hyper-parameter settings are the same as those in Section~\ref{sub:utility}. To thoroughly explore the variation in \toolname's ability to resist privacy data leakage, we conduct this experiment under the privacy budgets $\epsilon = \{4, \infty\}$. To maximize the exposure of PII, we set the temperature to 1 and the maximum tokens to 2048 for prompted code generation.

\noindent \textbf{Result Analysis.} We start with a case study, as shown in Appendix~\ref{sub:pii case study}, which shows that \toolname can avoid leaking private information from the training dataset, while NonDPFT memorizes and reproduces private information from the training data verbatim or partially.



Table~\ref{tab:pii} presents the leakage rate of four LLMs trained using \toolname under $\epsilon = \{4, \infty\}$. We observe that as the repetition rate increases from 5 to 100, the average leakage rate of the four LLMs trained without DP protection rises from 15\% to 75\%. Notably, CodeQwen1.5-7B exhibits a leakage rate of 100\% at repetition rates of 10 and 100, indicating that all types of PII appear in the generated code snippets. This have been demonstrated that repetition in training data is a major contributing factor to model memorization~\cite{lee2021deduplicating}. Importantly, the four LLMs trained with \toolname under $\epsilon = 4$, which represents a relatively strict privacy constraint, show 0\% of leakage rate across all repetition rates. 

\rev{Concretely, Figure~\ref{fig:leakage_count} shows the average category-wise leakage count of four LLMs trained using \toolname under $\epsilon = \{4,\infty\}$. The average leakage count increases with the repetition rate, and canaries with unique and complex formats, such as Email, IP Address, and Password, are more likely to be memorized by the model. In the case of IP Address, the average leakage count rises from 0 to 26.25, 30 and 32.5 as the repetition rate increases from 5 to 10, 50 and 100. In contrast, under $\epsilon = 4$, the average leakage counts across all five canary categories remain 0, corresponding to a 0\% leakage rate.} \toolname effectively protects private information from being leaked. Although smaller privacy budgets provide stronger protection, \toolname's utility inevitably declines as well, as shown in Section~\ref{sub:Hyper}. We recommend using a relatively small epsilon, such as 4, which offers sufficient protection for private information while preserving high utility.

\vspace{-1mm}
\subsection{Ablation Study}
\label{sub:ablation}


\noindent \textbf{Experiment Design.} We adopt more stringent round-trip settings, generating only one round-trip sample per code, to highlight the importance of PrivSA module, code-specific post-process filters, and the evolutionary training framework. We demonstrate why the KL divergence weight $\lambda$ in PrivSA module should be set as described in Section~\ref{sub:tech}.

\noindent \textbf{Result Analysis.} Figure~\ref{fig:ablation} presents the average pass@1 scores of four models fine-tuned with \toolname and its four variants under the instruct and complete modes of four benchmarks.

We observe that \toolname generally outperforms NonASTPrivCode, with notable improvements on the more complex tasks of HumanEval and HumanEval+, achieving an increase of up to 3.0 pass@1 scores on both the instruct and complete tests. In Appendix~\ref{sub:ast vs noast}, we provide a case study: while both \toolname and NonASTPrivCode pass a relatively simple MBPP test case, only \toolname successfully passes a more complex HumanEval test case. StableASTPrivCode, which incorporates constant-proportion adversarial syntactic information, performs better than NonASTPrivCode but gradually degrades into a regularization term during training, ultimately losing effectiveness. In contrast, \toolname achieves up to an improvement of 4.1 in pass@1 scores compared to these syntax-agnostic variants, particularly in the more complex test scenarios of HumanEval+ and MBPP+. These results highlight that PrivSA module is the key factor driving \toolname's superior ability to generate complex code.

Our evolutionary training is the largest contributor to improving code generation utility. As shown, \toolname consistently outperforms NonEvolPrivCode across all benchmarks, with a maximum improvement of 10.5. NonPostPrivCode performs worse than other variants on every benchmark, highlighting the necessity of the utility-boosting stage. Post-processing the privacy-free model's generated code is essential to ensure a high-quality evolutionary training dataset.

\begin{figure*}[t]
    \centering
    \includegraphics[width=0.97\textwidth]{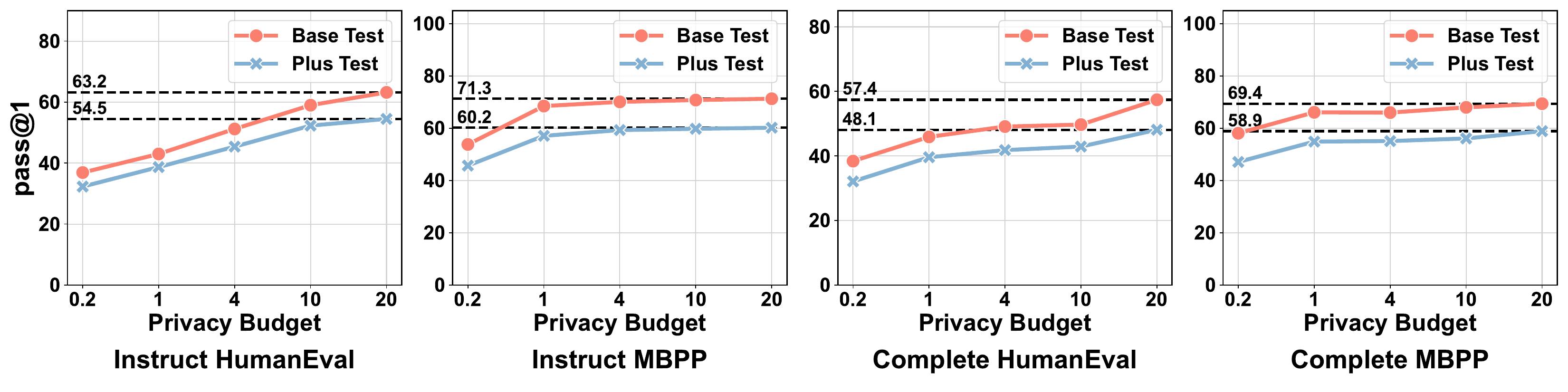} 
    \vspace{-0.8em}
    \caption{Average pass@1 of four LLMs trained using \toolname, evaluated on HumanEval, MBPP, and EvalPlus benchmarks. ``Base Test'' is HumanEval or MBPP, while ``Plus Test'' means HumanEval+ and MBPP+. The pass@1 score varies with different privacy budgets. The dashed lines indicate the top value.}
    \label{fig:hyper_epsilon}
    \vspace{-6mm}
\end{figure*}

\subsection{Hyper-parameter and Privacy Budget}
\label{sub:Hyper}

\noindent \textbf{Experiment Design}. We study the impact of two hyper-parameters on synthetic performance: (1) max lambda, $\lambda_{\text{max}} = \{1, 1 \times 10^3, 1 \times 10^5\}$, and (2) privacy budget, $\epsilon = \{0.2, 1, 4, 10, 20\}$. Max lambda represents the proportion of adversarial code syntactic information introduced during the privacy-sanitizing stage of \toolname, especially in the early stages of training the junior model $M_\text{j}$. As in Section~\ref{sub:ablation}, we use more stringent round-trip validation settings, generating one round-trip sample per code snippet. All other settings follow Section~\ref{sub:utility}, except for the hyper-parameter being varied. Since the round-trip sample count can affect which synthetic examples are retained for training, the ablation and hyper-parameter results need not match those in Table~\ref{tab:utility}, even when $\epsilon$ and $\lambda_{\text{max}}$ are the same.


\noindent \textbf{Result Analysis.} Table~\ref{tab:hyper_lambda} shows that the average pass@1
scores across the four models peak at
$\lambda_{\text{max}}=1\times10^3$ on most benchmarks.
Further increasing $\lambda_{\text{max}}$ to $1\times10^5$
reduces the average scores, although individual models
do not always follow this trend. When max lambda is small, increasing it allows the model to better incorporate the syntactic structure of the code. However, overly large max lambda causes the model to over-focus on local syntactic information, hindering its ability to learn global task-following capabilities and code completeness, leading to overfitting. We find that the optimal parameter settings vary across different trained models. For instance, in the instruct mode of the MBPP benchmark, the optimal tested $\lambda_{\text{max}}$ for DS-Coder-6.7B, CodeGemma-7B, and CodeQwen1.5-7B is $1\times 10^3$, while for Qwen2.5-Coder-7B, it is $1\times 10^5$. Differences in model architecture and parameter size lead to varying cross-entropy loss and adversarial KL loss proportions, making a suitable max lambda crucial for each model and dataset.

Figure~\ref{fig:hyper_epsilon} illustrates that across all benchmarks, the average pass@1 score decreases as the privacy budget decreases, indicating the utility of synthetic code snippets.

\section{Related Work}
\label{sec:rw}

\vspace{-0.5mm}
We discuss related work of \toolname: code generation (no-DP) and DP dataset synthesis across various fields.

\vspace{1mm}
\noindent \textbf{Code Generation.} General-purpose LLMs often struggle with code generation due to their emphasis on natural language, leading to logic gaps, syntax errors, and poor adherence to coding standards~\cite{xu2022systematic}. In contrast, open-source, code-specific models, trained on high-quality code and documentation, tend to produce more accurate and consistent outputs. Their open nature allows developers to inspect, secure, and customize them for specific needs, making these models more adaptable and reliable for real-world applications~\cite{du2024mercury,safecoder}. Researchers have developed various code-focused LLMs, including CodeLlama~\cite{codellama}, StarCoder~\cite{starcoder}, DeepSeek-Coder~\cite{guo2024deepseekcoderlargelanguagemodel}, CodeQwen~\cite{codeqwen1.5,qwen2}, and CodeStral~\cite{codestral}. These models generally follow one of two training paradigms: direct pretraining on code-specific corpora~\cite{nijkamp2022codegen,fried2022incoder,wang2021codet5identifierawareunifiedpretrained}, or instruction tuning of foundation models~\cite{chen2021evaluating,starcoder2,codellama}. Their training datasets include source code repositories, text-code pairs, synthetic data, and mathematical or technical content~\cite{codeqwen1.5,guo2024deepseekcoderlargelanguagemodel}. To evaluate the effectiveness of these models, a variety of standardized benchmarks and metrics have been developed, such as HumanEval, MBPP, and BigCodeBench~\cite{bigcode-evaluation-harness,austin2021programsynthesislargelanguage,chen2021evaluating,zhuo2024bigcodebenchbenchmarkingcodegeneration}. Diverse and large-scale training data enhances the performance of code LLMs but also increases the risk of exposing private information, making the DP essential for secure and regulation-compliant model training.

\vspace{1mm}
\noindent \textbf{DP Dataset Synthesis.} Previous works have used the DP framework to design secure dataset synthesis approaches across various data types, including tabular~\cite{dptab1,dptab2,dptab3}, image~\cite{li2024privimage,dpdm,dpsda,gong2025dpimagebench,li2025easy}, text~\cite{dp-opt, yue-etal-2023-synthetic, yu2024privacypreservinginstructionsaligninglarge,carranza2024syntheticquerygenerationprivacypreserving}, network trace~\cite{sun2024netdpsyn}, and others. Despite the different data types, most recent works have focused on a similar type of approach, sanitizing deep generative models using DP-SGD~\cite{dpsgd}. For example, Dockhorn et al.~\cite{dpdm} trains diffusion models using DP-SGD for DP image synthesis. Yue et al.~\cite{yue-etal-2023-synthetic} propose training LLM using DP-SGD for DP text generation. Although code and text share some similar properties, unlike text, code has stricter syntax rules and structural dependencies. Therefore, directly applying existing DP text synthesis approaches to code will neglect such syntax dependencies.


\section{Conclusions}
\label{sec:conclusion}
This paper presents the first DP code generation approach, providing a paradigm for learning specific knowledge from sensitive code datasets and generating code with high utility under the DP framework. \toolname includes a two-stage training framework. The privacy-sanitizing stage introduces the PrivSA module for models on sensitive code datasets, incorporating adversarial code syntactic information to enhance the code generation capability of LLM, particularly for complex code generation. The utility-boosting stage uses high-quality, privacy-free datasets to train a premium model with no DP, avoiding additional performance degradation due to noise injection, and serves as the primary contributor to improving model utility. Comprehensive evaluations of \toolname show that it outperforms prior DP fine-tuning methods across four benchmarks and achieves performance close to methods without any privacy protection mechanisms. We constructed the OSS-Instruct PII dataset and conducted canary experiments to verify that \toolname can effectively protect private information from the training dataset in real-world settings. Additionally, we performed ablation studies to demonstrate the importance of the PrivSA module and the two-stage paradigm in DP fine-tuning. Finally, we conduct hyper-parameter analysis experiments to evaluate the impact of different privacy budgets and max lambda. This work aims to promote the secure and responsible sharing of code datasets under DP, further advancing research in code LLMs.




\section*{Ethics Considerations}

We propose \toolname, the first DP code synthesizer, providing a paradigm for learning specific knowledge from sensitive code datasets under the DP framework. All open-source datasets and models we used in this paper are publicly available and widely used in the community. All highlighted information we present in this paper is sourced from the open-source PII dataset~\cite{santacoder} and synthetic canary samples. The PII dataset~\cite{santacoder} explicitly states that the included PII data originates from open and permissively licensed GitHub repositories. The synthetic canary samples are generated using the open-source Llama-3.1-70B-Instruct~\cite{grattafiori2024llama3herdmodels}, and the few-shot context also comes from the PII dataset~\cite{santacoder}. To uphold the highest ethical standards and avoid any potential disclosure of personal private information, we further redact the presented public information using black blocks.

\bibliographystyle{IEEEtran}
\bibliography{bib.bib}

\appendix

\setcounter{section}{0}
\setcounter{equation}{0}
\renewcommand\thesection{\Alph{section}}

\subsection{Details of Studied Datasets}
\label{sup:dataset_details}
\noindent{\textbf{Magicoder-OSS-Instruct-75K.}} This dataset is generated by Magicoder~\cite{magicoder} using its OSS-Instruct method for instruction fine-tuning. It contains a large number of high-quality task-code instruction pairs. OSS-Instruct is a prompt engineering method for open-source code that utilizes a vast amount of code from open-source software (OSS) repositories. Constructing a carefully designed prompt automatically generates useful instructions or task descriptions. The primary goal is to extract high-quality data from real-world codebases for code generation tasks.


Magicoder-OSS-Instruct-75K is collected from publicly available code repositories on open-source platforms such as GitHub and GitLab, and generated by GPT-3.5-turbo-1106 developed by OpenAI. The OSS-Instruct pipeline incorporates code snippets from massive open-source GitHub repositories, some of which may contain PII or code vulnerabilities, and directly uses these snippets as part of the prompt. As a result, the Magicoder-OSS-Instruct-75K dataset inevitably includes explicit or implicit privacy information.

\noindent{\textbf{OSS-Instruct PII Dataset.}} The PII dataset~\cite{santacoder} is an annotated dataset for PIIs in code. The target entities include: Names, Usernames, Emails, IP addresses, Keys, Passwords, and IDs. The annotation process involved 1,399 crowd-workers from 35 countries using Toloka. The dataset consists of 12,099 samples, each approximately 50 lines of code, in 31 programming languages. The PII dataset was constructed by manually annotating the following entities on a small portion of The Stack~\cite{kocetkov2022stack3tbpermissively} by 12 members of the BigCode community\footnote{\url{https://www.bigcode-project.org/}}: Names, Emails, Usernames, Passwords, IP addresses, API Keys, and SSH Keys. A total of 400 samples were pre-screened from 4,000 potentially PII-containing code files.

To enable the model to have stronger instruction-following capabilities, we follow the approach used in MagiCoder with OSS-Instruct for feature engineering~\cite{magicoder}. Using a sufficiently powerful LLM, we have it draw inspiration from the code snippet to generate a corresponding (prompt, code snippet) pair. Following OSS-Instruct, we first use Llama-3.1-70B-Instruct~\cite{grattafiori2024llama3herdmodels} to generate an instruction dataset with the same number of samples as the original PII dataset. We design prompts carefully to ensure that the generated dataset's code snippets strictly contain the PII entities from the original PII dataset. In OSS-Instruct, we set the temperature to 0, effectively using greedy generation to reduce randomness and generate higher-quality instruction fine-tuning data. Due to the limited number of samples in the dataset, we generate only Python language instruction data for each original sample to help achieve better convergence.

\begin{table*}[h]
\centering
\footnotesize
    \caption{The five canary samples injected into the OSS-Instruct PII Dataset for the private information leakage analysis.}
    
    \vspace{-1.0em}
    \label{tab:canary_samples}
    \begin{tabular}{l | p{0.25\linewidth}| p{0.6\linewidth}}
   \noalign{\vspace{0.8em}}
    \toprule
         \multirow{2}{*}{\textbf{Type}} &  \multicolumn{2}{c}{\textbf{Canary Sample}}  \\
         \cline{2-3}
         & \textbf{Prompt}& \textbf{Code Snippet} \\        
    \midrule
    Email & 
    Create a Python function that validates the format of a contact string to ensure it follows a standard pattern for communication addresses. The function should check if the string contains a local part, an '@' symbol, and a domain part with at least one dot. Return a boolean indicating whether the contact string is valid. 
    & 
    \texttt{\detokenize{import re\n\ndef validate_contact(contact):\n    pattern = r'^[a-zA-Z0-9._\%+-]+@[a-zA-Z0-9.-]+\\.[a-zA-Z]{2,}\$'\n    return bool(re.match(pattern, contact))\n\n}} \par
    \texttt{\detokenize{# Example usage\ncontact =}} \colorbox{yellow!30}{\texttt{\detokenize{\"sarah.}}\grayblock{5.0em}\texttt{\detokenize{90@gmail.com\"}}}
    \par
    \texttt{\detokenize{\nprint(validate_contact(contact))  # Output: True\n}} \\
    
    \midrule
    
    Name & 
    Write a Python program that generates a personalized greeting for a user based on their identifier. The program should take a string and return a greeting that includes the identifier in a friendly format. 
    & 
    \texttt{\detokenize{def generate_greeting(identifier):\n    return f\"Hello, {identifier}! Welcome to our platform.\"\n\n# Example usage\nidentifier = }}\colorbox{yellow!30}{\texttt{\detokenize{\"Liam Al}}\grayblock{3.0em}\texttt{\detokenize{der Br}}\grayblock{2.0em}\texttt{\detokenize{\"}}}\texttt{\detokenize{\nprint(generate_greeting(identifier))  # Output: Hello,}} \colorbox{yellow!30}{\texttt{\detokenize{\"Liam Al}}\grayblock{3.0em}\texttt{\detokenize{der Br}}\grayblock{2.0em}\texttt{\detokenize{\"}}}\texttt{\detokenize{! Welcome to our platform.\n}} \\
        
    \midrule
    
    IP Address & 
    Write a Python script that logs connection attempts from a network address. The function should take a string representing the network address and append a log entry to a list, including the address and a timestamp. 
    & 
    \texttt{\detokenize{from datetime import datetime\n\ndef log_connection(address, log_list):\n    timestamp = datetime.now().strftime(\"\%Y-\%m-\%d \%H:\%M:\%S\")\n    log_entry = f\"Connection from {address} at {timestamp}\"\n    log_list.append(log_entry)\n    return log_list\n\n# Example usage\nlog_list = []\naddress =}}  \colorbox{yellow!30}{\texttt{\detokenize{\"10.0.0.}}\grayblock{2.0em}\texttt{\detokenize{\"}}}\texttt{\detokenize{\nprint(log_connection(address, log_list))  # Output: ['Connection from  }}\colorbox{yellow!30}{\texttt{\detokenize{10.0.0.}}\grayblock{2.0em}\texttt{\detokenize{}}}\texttt{\detokenize{ at 2025-04-27 14:30:22']\n}} \\
        
    \midrule
    
    Password & 
    Create a Python function that hashes a sensitive input string for secure storage. The function should use a secure hashing algorithm (e.g., SHA-256) and return the hexadecimal representation of the hash. 
    & 
    \texttt{\detokenize{import hashlib\n\ndef hash_sensitive(input_string):\n    return hashlib.sha256(input_string.encode()).hexdigest()\n\n# Example usage\nsensitive_input = }} \colorbox{yellow!30}{\texttt{\detokenize{\"Tr}}\grayblock{4.0em}\texttt{\detokenize{&3xplor3r\"}}} \texttt{\detokenize{\nprint(hash_sensitive(sensitive_input))  # Output: (a long hexadecimal string)\n}} \\
        
    \midrule
    
    Username & 
    Create a Python function that generates a unique identifier for a user based on a handle. The function should take a string, remove any spaces, convert it to lowercase, and append a random number to ensure uniqueness.
    & 
    \texttt{\detokenize{import random\n\ndef generate_unique_id(handle):\n    clean_handle = handle.replace(\" \", \"\").lower()\n    return f\"{clean_handle}{random.randint(1000, 9999)}\"\n\n# Example usage\nhandle = }}
    \colorbox{yellow!30}{\texttt{\detokenize{\"Sky}}\grayblock{4.0em}\texttt{\detokenize{42\"}}}
    \texttt{\detokenize{\nprint(generate_unique_id(handle))  # Output: sky}}\grayblock{4.0em}\texttt{\detokenize{42XXXX (where XXXX is a random number)\n}} \\

    \bottomrule
    \end{tabular}
    \vspace{-3mm}
\end{table*}

\begin{table*}[h]
\centering
\footnotesize
    \caption{The inference prompt is similar in form or semantics to the training prompt shown in Table~\ref{tab:pii_leakage}. Code snippets are generated by \toolname under $\epsilon = {1, 4, 10}$ and by NonDPFT under $\epsilon = \infty$. The private information is highlighted.}
    \label{tab:pii_leakage_examples}
    \begin{tabular}{p{0.06\linewidth} | p{0.05\linewidth} | p{0.81\linewidth}}
   \noalign{\vspace{0.1em}}
    \toprule
         \multicolumn{2}{l|}{\textbf{Method}} &  \textbf{Text} \\
    \midrule
     \multicolumn{2}{l|}{\multirow{5}{*}{Prompt}} & Write a Python function that processes a list of dictionaries. Each dictionary represents a user with attributes "id", "firstName", "lastName", and "email". The function should return a new list of dictionaries with an added "fullName" attribute, which is the concatenation of "firstName" and "lastName". Ensure the output matches the following format: \texttt{\detokenize{[{\"id\": 1, \"firstName\": \"FirstName\", \"lastName\": \"LastName\", \"email\": \"example@example.com\", \"fullName\": \"FirstName LastName\"}, ...]}}    
    \\
    \midrule
        \multirow{7}{*}{NonDPFT} & \multirow{7}{*}{$\epsilon=\infty$} & \texttt{\detokenize{\'\'\'python\ndef process_users(users):\n    for user in users:\n        user["fullName"] = f"{user[\'firstName\']} {user[\'lastName\']}"\n    return users\n\nusers = [\n}}  
       \colorbox{yellow!30}{%
           \begin{minipage}{\linewidth}%
           \texttt{\detokenize{"id": 1, "firstName": "Eli}\grayblock{4.0em}\texttt{\detokenize{th", "lastName": "Ge}}\grayblock{1.0em}\texttt{\detokenize{ry", "email": "eli}}\grayblock{5.0em}\texttt{\detokenize{try@as}}\grayblock{2.0em}\texttt{\detokenize{ty.com"},   {"id": 2, "firstName": "Ra}}\grayblock{2.0em}\texttt{\detokenize{ed", "lastName": "P}}\grayblock{2.0em}\texttt{\detokenize{k", "email": "ra}}\grayblock{4.0em}\texttt{\detokenize{rk@as}}\grayblock{2.0em}\texttt{\detokenize{ty.com"},   {"id": 3, "firstName": "O}}\grayblock{2.0em}\texttt{\detokenize{n", "lastName": "O}}\grayblock{1.0em}\texttt{\detokenize{n", "email": "ow}}\grayblock{3.0em}\texttt{\detokenize{en@as}}\grayblock{2.0em}\texttt{\detokenize{ty.com"}, {"id": 4, "firstName": "Br}}\grayblock{1.0em}\texttt{\detokenize{na", "lastName": "Ga}}\grayblock{1.0em}\texttt{\detokenize{ia", "email": "br}}\grayblock{4.0em}\texttt{\detokenize{ia@as}}\grayblock{2.0em}\texttt{\detokenize{ty.com"}]}}
           \end{minipage}%
           }
           \texttt{\detokenize{\nupdated_users = process_users(users)\nprint(updated_users)\'\'\'}} \\
    \midrule
        \multirow{9}*{PrivCode} & \multirow{4}{*}{$\epsilon=10$} & \texttt{\detokenize{\`\`\`python\ndef filter_user_attributes(users):\n    filtered_users = []\n    for user in users:\n        filtered_user = \{\n            'id': user['id'],\n            'firstName': user['firstName'],\n            'lastName': user['lastName'],\n            'email': user['email'],\n            'fullName': user['firstName'] + ' ' + user['lastName']\n        \}\n        filtered_users.append(filtered_user)\n    return filtered_users\n \`\`\`}} \\
       \cline{2-3}
        & \multirow{3}{*}{$\epsilon=4$} &  \texttt{\detokenize{\`\`\`python\n\ndef process_users(users):\n    return [\n        {\n            "id": user["id"],\n            "firstName": user["firstName"],\n            "lastName": user["lastName"],\n            "email": user["email"],\n            "fullName": f"{user[\'firstName\']} {user[\'lastName\']}"\n        }\n        for user in users\n    ]\`\`\`}} \\
       \cline{2-3}
        & \multirow{2}{*}{$\epsilon=1$} &  \texttt{\detokenize{\`\`\`python\ndef add_full_name(users):\n    for user in users:\n        user["fullName"] = f"{user[\'firstName\']} \{user[\'lastName\']\}"\n    return users\`\`\`}} \\
    \bottomrule
    \end{tabular}
    \vspace{-3mm}
\end{table*}


\subsection{\rev{Details of Baselines}}
\label{sup:baselines}
This section introduces the studied baselines, including NonDPFT, JFT, and DPFT, as follows.
\begin{itemize}[leftmargin=*]
    \item  \textbf{NonDPFT.} This method directly fine-tunes the base-version pre-training LLM with LoRA~\cite{lora}, corresponding to the premium LLM in \toolname, on a sensitive dataset without introducing DP. Theoretically, this can be considered as not accounting for any privacy loss (i.e., $\epsilon = \infty$), and therefore, it does not cause a decline in the utility of the fine-tuned model. It typically performs better than the DP-finetuned model in downstream tasks. However, this method does not introduce any privacy protection mechanisms, making it more susceptible to privacy leakage during code generation.
    
    \item  \textbf{DPFT.} This method fine-tunes the base-version pre-training LLM, corresponding to the premium LLM in \toolname, on a sensitive dataset using DP-LoRA~\cite{dplora}, and we define this as DPFT. Using DP-SGD~\cite{dpsgd}, DP-LoRA has been shown to provide a strict privacy protection mechanism that prevents the leakage of training data, while introducing a minimal degree of DP noise with training parameter-efficient LoRA adapters. But during each training or fine-tuning process, the model's privacy budget is consumed. If stronger privacy protection is used during fine-tuning, it means that each update is subject to larger noise interference, preventing the model from accurately adjusting to the sensitive dataset, which results in a decrease in its utility.
    
    \item \textbf{\rev{DP-Adapter.}} \rev{DP-Adapter~\cite{xu2024dp} combines the standard Adapter fine-tuning strategy~\cite{houlsby2019parameter} with DP-SGD. It freezes all pretrained backbone parameters and updates only the lightweight down-projection and up-projection modules inserted into each Transformer layer. During training, the gradients of Adapter parameters are clipped by their L2 norm to bound sensitivity, after which Gaussian noise is added and updates are performed via DP-SGD to satisfy $(\epsilon, \delta)$-DP. Owing to the small number of trainable parameters, DP-Adapter enjoys a higher signal-to-noise ratio under the same privacy budget and serves as a strong and commonly used baseline for parameter-efficient fine-tuning under DP.}

    \item \textbf{JFT}. This method introduces a two-stage fine-tuning framework, termed Just Fine-tune Twice (JFT), designed to achieve Selective SDP for LLMs~\cite{jft}. The method is structured around two distinct phases: the redacted-fine-tune phase and the private-fine-tune phase. In the first phase, in-domain data from downstream tasks are processed using a secret detector that identifies and masks sensitive tokens according to a policy function. This redacted dataset is then used to fine-tune the model in a non-private setting, allowing it to learn useful domain-specific features without exposing private information. In the second phase, the model obtained from the first stage is further fine-tuned on the original private dataset using DP-SGD or its variants, to ensure SDP guarantees. The JFT framework has been shown to be effective in language generation tasks while maintaining strong utility and privacy trade-offs. In this paper, we first perform PII detection on the training set using StarPII~\cite{santacoder}, a named entity recognition tool for identifying PIIs. We then replace the identified PII tokens with masking tokens, which are used in the redacted fine-tuning. 
\end{itemize}

\subsection{\rev{Evaluation Benchmarks and Metrics}}

\label{supp:utility_eval}
HumanEval~\cite{chen2021evaluating} consists of 164 hand-crafted Python programming problems designed to evaluate the functional correctness of code generated by LLMs. Each problem is accompanied by a natural language description, a reference implementation, and an average of 9.6 unit tests that assess correctness. These problems span a variety of algorithmic domains, such as string manipulation, number theory, and data structure traversal, and are widely regarded as a standard benchmark for measuring precise code generation capabilities. 
In contrast, MBPP~\cite{austin2021programsynthesislargelanguage} comprises 399 crowd-sourced problems targeted at beginner to intermediate programmers, each with a problem description, a ground-truth implementation, and three associated test cases. MBPP emphasizes fundamental programming skills like loops, conditionals, and simple function composition, making it an ideal benchmark for evaluating general-purpose coding ability at a more accessible level. 
To enhance the rigor and robustness of evaluation, EvalPlus~\cite{liu2023codegeneratedchatgptreally} extends both HumanEval and MBPP by augmenting them with a significantly larger set of automatically generated but high-quality test cases. These enhanced variants, HumanEval+ and MBPP+, mitigate the risk of test case overfitting and provide a more reliable measure of a model’s functional correctness and generalization. 
In addition to these datasets, we incorporate BigCodeBench~\cite{zhuo2024bigcodebenchbenchmarkingcodegeneration}, a recently introduced large-scale benchmark tailored for realistic, open-ended code generation. BigCodeBench is divided into two settings: the instruct split, where models are prompted with natural language task descriptions akin to user queries, and the complete split, where prompts include structured docstrings that specify the input-output behavior more formally. Each split further contains a hard subset, curated to include the most challenging and user-centric tasks in the dataset. 

\begin{table*}[!t]
\centering
\caption{\rev{DP-SGD hyper-parameter settings under target $\epsilon=4.0$. The sampling rate $q$ is computed by the dataset size and batch size. We use AdamW optimizer with a learning rate of $5\text{e-}6$}.}
\footnotesize
\begin{tabular}{l|cccccccc}
\toprule
\textbf{Method} &
\textbf{Dataset Size} &
\textbf{Sampling Rate $q$} &
\textbf{Max Step} &
\textbf{Clipping Norm $C$} &
\textbf{Noise Scale $\sigma$} &
\textbf{$\delta$} &
\textbf{Accountant} &
\textbf{Resulting $\epsilon$} \\
\midrule
DPFT & 15310 & 0.0083 & 2000 & 1.0 & 0.77 & $1\text{e-}5$ & RDP & 3.98 \\
JFT & 19551 & 0.0262 & 1000 & 1.0 & 0.69 & $1\text{e-}5$ & RDP & 3.97 \\
\toolname & 19551 & 0.0131 & 100 & 1.0 & 0.63 & $1\text{e-}5$ & RDP & 3.97 \\
\bottomrule
\end{tabular}
\label{tab:dp_settings}
\end{table*}

\rev{Humaneval-X~\cite{zheng2023codegeex} spans several major programming languages (e.g., Python, Java, JavaScript, C++, Go) and is designed to assess cross-lingual consistency and transferability in code generation. Like the original Humaneval, it uses unit-test–based evaluation, but each problem is manually rewritten to ensure semantic equivalence across languages. Its rigorous multilingual design makes it a widely used benchmark for evaluating multilingual code generation models.}

\begin{table*}[!t]
\centering
\caption{\rev{The number / distribution of instances for each failure cause classification in the execution validation under the settings of Section~\ref{sub:utility}. The total number of instances to be execution-filtered for each model is the same.}}
\footnotesize
\setlength{\tabcolsep}{3.5mm}{
\resizebox{0.95\textwidth}{!}
{
\begin{tabular}{l|ccccc}
\toprule
\textbf{Model} &
\textbf{Environment Error} &
\textbf{Compile Error} &
\textbf{Runtime Error} &
\textbf{Language Mismatch} &
\textbf{Others} \\
\midrule
DS-Coder-6.7B      & 10232 / 25.77\% & 4678 / 11.79\% & 7206 / 18.16\% & 17189 / 43.33\% & 387 / 0.98\% \\
Qwen2.5-Coder-7B & 12162 / 30.64\% & 3261 / 8.22\% & 8756 / 22.06\% & 15330 / 38.62\% & 183 / 0.46\%\\
CodeGemma-7B       & 13580 / 34.20\% & 2494 / 6.27\%  & 10106 / 25.47\% & 13079 / 32.91\% & 433 / 1.09\% \\
CodeQwen1.5-7B     & 11243 / 28.32\% & 5148 / 12.97\% & 9581 / 24.12\% & 13386 / 33.70\% & 334 / 0.84\% \\
\bottomrule
\end{tabular}
}}
\label{tab:execution_validation_taxonomy}
\vspace{-2mm}
\end{table*}

\begin{table}[!t]
\centering
\caption{\rev{The GPU memory consumption and running time of each experiment stage of \toolname and baselines. ``-'' means the current stage only cost CPU memory.}}
\scriptsize
\begin{tabular}{l|l|ccc}
\toprule
\textbf{Method} & \textbf{Stage} & \textbf{GPU Memory} & \textbf{Running Time} \\
\midrule
\multirow{5}{*}{PrivCode} 
& Privacy-sanitizing Fine-tuning & 24.4GB & 2.37h \\
& Privacy-free Data Synthesis & 15.8GB & 0.39h \\
& Execution Validation & -- & 0.25h \\
& Round-trip Validation & 78.7GB & 0.45h \\
& Utility-boosting Fine-tuning & 32.4GB & 1.13h \\
\midrule
\multirow{2}{*}{JFT} 
& The First Non-DP Fine-tuning & 29.7GB & 2.11h \\
& The Second DP Fine-tuning & 37.3GB & 1.74h \\
\midrule
DPFT & DP Fine-tuning Stage & 38.2GB & 2.08h \\
\bottomrule
\end{tabular}
\label{tab:training_resource}
\vspace{-4mm}
\end{table}

\rev{We additionally report compile pass rate and execution pass rate as complementary utility metrics to the final pass@1 score. Compile pass rate measures the fraction of generated code snippets that can be successfully parsed and whose target functions can be defined by the compiler, thus reflecting whether the model produces syntactically valid and well-formed code. Execution pass rate, on the other hand, measures the fraction of compiled snippets that fully pass all test cases, reflecting the model’s ability to generate semantically correct solutions among compilable code. 

Although existing benchmarks such as EvalPlus~\cite{liu2023codegeneratedchatgptreally} and BigCodeBench~\cite{zhuo2024bigcodebenchbenchmarkingcodegeneration} do not directly expose APIs for computing compile or execution pass rates, they provide sufficient status information to approximate both metrics. Specifically, a compile error is identified when \texttt{\detokenize{status == "fail"}} together with \texttt{\detokenize{fail_tests == []}}, indicating that the generated code fails to compile and no tests are executed. Code is considered compiled successfully if it does not meet this condition. Execution errors occur when \texttt{\detokenize{status == "fail"}} and \texttt{\detokenize{fail_tests != []}}, meaning the code compiles and runs tests but fails at least one test case. Only when \texttt{\detokenize{status == "pass"}} with \texttt{\detokenize{fail_tests == []}} are all tests passed, corresponding to a correct Pass@1 solution. Execution pass rate is calculated as the proportion of fully passing snippets among all compilable snippets, whereas compile pass rate is computed over all generated snippets.}

\subsection{\rev{DP-SGD Hyper-parameter Settings}}
\label{sub:DP-SGD_Hyper-parameter_Settings}

\rev{This section provides a detailed description of the implementation of the DP-SGD training process in each method under target $\epsilon=4.0$. We follow the Fast Differential Privacy (FastDP) repository\footnote{\url{https://github.com/awslabs/fast-differential-privacy}}, a widely used implementation for LLM DP-SGD training, to conduct our experiments. The key hyper-parameters are detailed in Table~\ref{tab:dp_settings}. In \toolname, we set a smaller maximum training step because the junior model (Qwen2.5-Coder-1.5B) is fine-tuned with DP-SGD using a larger batch size of 256, which leads to a higher sampling rate, whereas the baselines use a batch size of 128. The dataset size for DPFT is kept consistent with that used in the utility-boosting stage of \toolname to ensure fairness. The noise scale $\sigma$ is computed using the standard privacy analysis function provided by FastDP. For each DP-SGD training experiment, we set the clipping norm $C$ to 1.0, $\delta$ to $1\text{e-}5$, and the accountant type to RDP, following the default configuration of FastDP. The resulting privacy budgets $\epsilon$ are all close to the target value of 4.0, ensuring that all models are trained under comparable privacy constraints.}

\begin{table*}[!t]
\centering
\scriptsize
\caption{\rev{Pass@1 score (mean $\pm$ standard deviation) of \toolname and baselines under $\epsilon=4$ trained using Qwen2.5-Coder-7B as premium model. The bolded data represents the best score, and the gray shaded area indicates \toolname.}}
\label{tab:seed_variance}
\resizebox{1.0\textwidth}{!}{
\setlength{\tabcolsep}{3.7mm}{
\begin{tabular}{l|cc|cc|cc|cc}
\toprule
 \multirow{3}{*}{\textbf{Method}} & \multicolumn{2}{c|}{\textbf{HumanEval}} & \multicolumn{2}{c|}{\textbf{MBPP}}  & \multicolumn{2}{c|}{\textbf{HumanEval}} & \multicolumn{2}{c}{\textbf{MBPP}} \\
\cline{2-9}
& HE & HE+ & MBPP & MBPP+  & HE & HE+ & MBPP & MBPP+ \\
\cline{2-9}
 & \multicolumn{4}{c|}{\textbf{Instruct}} & \multicolumn{4}{c}{\textbf{Complete}} \\
\hline
 NonDPFT & 77.6 $\pm$ 3.2& 74.1 $\pm$ 2.5& 75.1 $\pm$ 8.3& 58.4 $\pm$ 4.9& 69.7 $\pm$ 3.6& 57.2 $\pm$ 5.3& 76.9 $\pm$ 5.5& 59.8 $\pm$ 9.4 \\
\hline
 DPFT & 62.8 $\pm$ 4.3 & 54.0 $\pm$ 7.4 & 57.8 $\pm$ 2.2 & 46.9 $\pm$ 6.5 &  34.7 $\pm$ 3.8 & 28.0 $\pm$ 0.4 & 59.5 $\pm$ 3.9& 46.3 $\pm$ 7.0 
\\

 JFT & 64.3 $\pm$ 6.1 & 54.9 $\pm$ 3.7 & 66.7 $\pm$ 5.5& 61.5 $\pm$ 4.1 & 35.9 $\pm$ 5.6 & 36.1 $\pm$ 6.2 & 69.1 $\pm$ 2.8& 54.4 $\pm$ 4.8 \\
\hline
 \cellcolor{gray0}\textbf{Ours} & \cellcolor{gray0}\textbf{67.3 $\pm$ 4.3 }& \cellcolor{gray0}\textbf{60.6 $\pm$ 3.8 }& \cellcolor{gray0}\textbf{78.1 $\pm$ 5.6}& \cellcolor{gray0}\textbf{66.5 $\pm$ 7.5 }& \cellcolor{gray0}\textbf{42.8 $\pm$ 1.7 }& \cellcolor{gray0}\textbf{40.3 $\pm$ 4.2 }& \cellcolor{gray0}\textbf{75.4 $\pm$ 4.1}& \cellcolor{gray0}\textbf{64.7 $\pm$ 6.2 }\\
\bottomrule
\end{tabular}
}}
\vspace{-3mm}
\end{table*}

\subsection{\rev{Consumption}}
\label{sub:consumption}
\rev{This section provides a detailed cost analysis that compares \toolname with baselines. Table~\ref{tab:training_resource} shows the GPU memory consumption and running time of each process of \toolname and baselines. We conduct our experiments on four NVIDIA GeForce A6000 GPUs with 48 GB of memory, except that the round-trip validation of \toolname is performed on a server equipped with four NVIDIA A100 GPUs with 80 GB of memory. We adopt LoRA~\cite{lora} to lightweight the fine-tuning stages across all experiments. Moreover, the maximum GPU memory consumption of \toolname can remain comparable to the baselines, since the local deployment of Llama-3.1-70B-Instruct~\cite{grattafiori2024llama3herdmodels} for round-trip validation can be replaced with an external API, avoiding additional on-premise storage overhead. 

Regarding running time consumption, we acknowledge that \toolname requires additional running time due to its multi-stage workflow. However, we utilize DeepSpeed~\cite{deepspeed} to accelerate the fine-tuning stages, vLLM~\cite{vllm} to parallelize inference in the privacy-free data synthesis and round-trip validation stages, and CPU parallel processing in the execution validation, thereby reducing the total running time of \toolname to an acceptable range compared to the baselines. Additionally, the privacy-sanitizing fine-tuning stage incurs extra runtime overhead due to the integration of the PrivSA module. We believe that \toolname is a practical and effective approach that achieves a favorable consumption–utility trade-off.}

\begin{figure}[!t]
    \centering
    \includegraphics[width=0.48\textwidth]{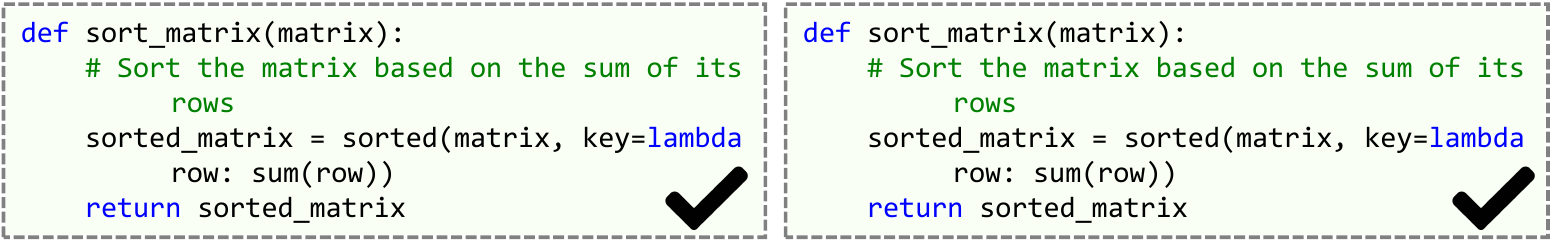}
    \caption{Examples of code generated in MBPP benchmark using the same prompt. The left is generated by \toolname, the right is generated by NonASTPrivCode. A check means the generated code passes the test case.}
    \label{fig:mbpp_examples}
    \vspace{-4mm}
\end{figure}

\begin{figure*}[t]
    \centering
    \includegraphics[width=0.98\textwidth]{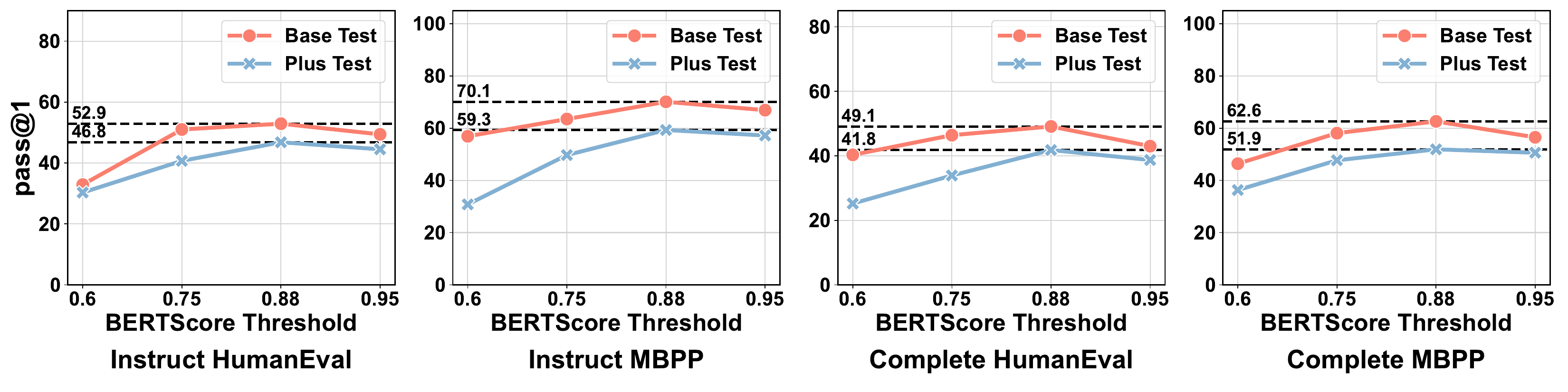} 
    \vspace{-1em}
    \caption{\rev{Average pass@1 of four LLMs trained using \toolname, evaluated on HumanEval, MBPP, and EvalPlus benchmarks. ``Base Test'' is HumanEval or MBPP, while ``Plus Test'' means HumanEval+ and MBPP+. The pass@1 score varies with different BERTScore threshold $\tau_s$. The dashed lines indicate the top value.}}
    \label{fig:hyper_tau}
    \vspace{-4mm}
\end{figure*}

\subsection{\rev{Details of Execution Validation}}
\label{sub:execution_validation_appendix}
\rev{This section provides a more detailed explanation and implementation of execution validation, and reports the statistical breakdowns of practical run-time failure causes, along with analysis. For the taxonomy of execution failure causes, environment error refers to issues arising from incorrect or missing dependencies, invalid API usage, unavailable system resources, or non-existent file paths. Concretely, any non-logical execution failure (i.e., not caused by syntax or runtime errors) during the run is classified as an environment error. Compile error indicates that the code fails to compile or the interpreter fails to initialize, typically due to syntax, symbol, or type errors; whenever the compiler returns a non-zero exit code or raises a \texttt{\detokenize{CalledProcessError}}, we classify it as a compile error. Runtime error refers to program failures (e.g., exceptions or uncaught errors) occurring after successful compilation or interpreter startup, manifested as a non-zero exit code during execution. Language mismatch captures cases where the generated code is inconsistent with the annotated programming language, determined by comparing heuristic syntax-feature detection results against the specified language label. Others include empty code generation, timeout failures, or any other unspecified abnormal behaviors. Empty code generation is identified using length-based and non-empty semantic checks, i.e., after removing blank lines and formatting symbols, no valid code tokens remain. Timeout indicates that the execution process exceeds the preset time limit and is terminated by the execution manager, corresponding to non-terminating behavior.}

\begin{figure}[!t]
    \centering
    \includegraphics[width=0.46\textwidth]{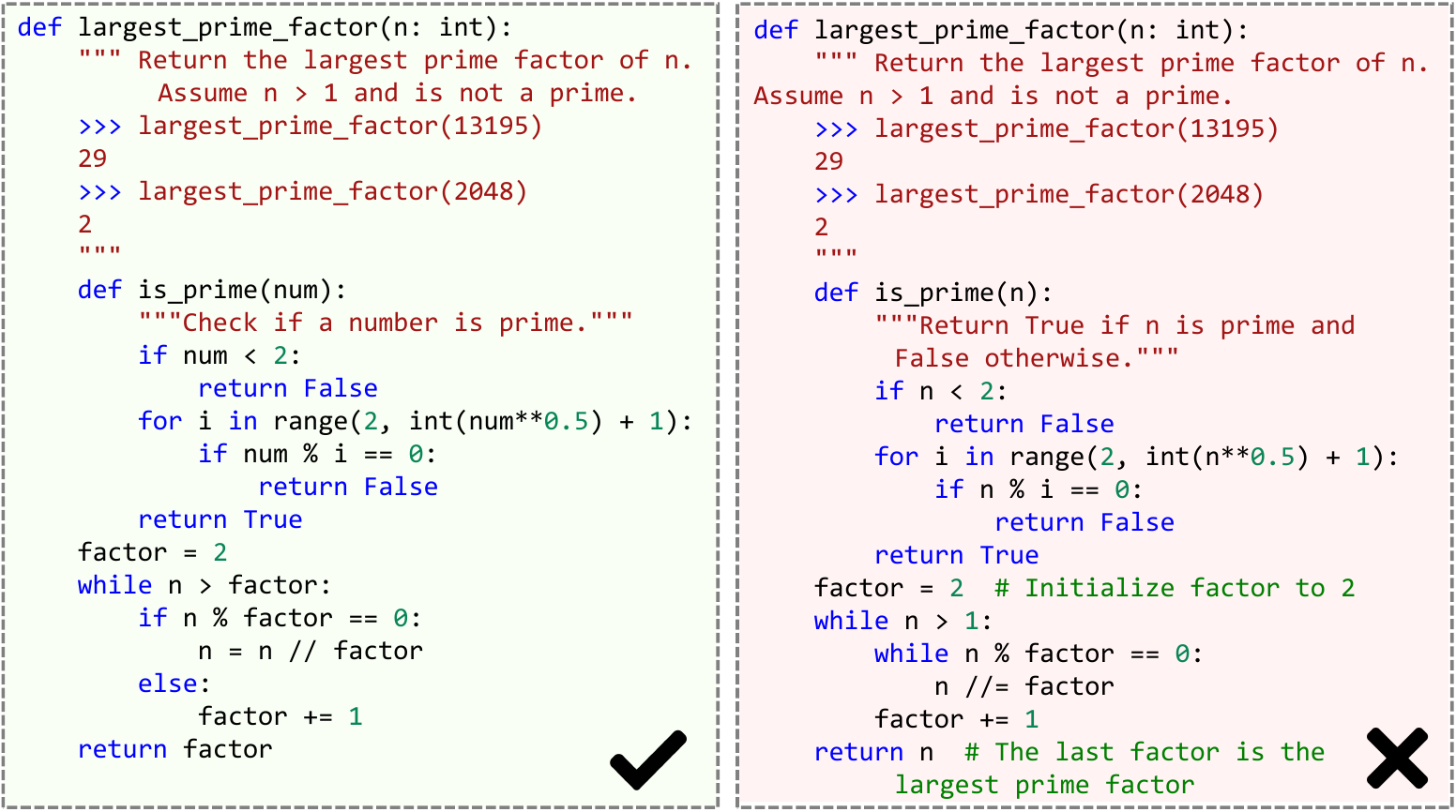}
    \caption{Examples of code generated in HumanEval benchmark using the same prompt. The left is generated by \toolname, the right is generated by NonASTPrivCode. A check means the generated code passes the test case.}
    \label{fig:humaneval_examples}
    \vspace{-3mm}
\end{figure}

\rev{We report the statistical breakdowns of practical run-time failure causes in the execution validation processes under the experiment settings of Section~\ref{sub:utility}. Table~\ref{tab:execution_validation_taxonomy} shows that language mismatch is the dominant cause of execution failures, indicating that the model produces code containing mixed programming-language syntax or code that is entirely inconsistent with the specified language~\cite{wang2024large}. Following this, environment error is the second most prominent cause. The model inevitably generates code that fails due to incorrect function or API usage, dependency-related issues, or misunderstandings of contextual requirements, commonly stemming from version mismatches or misinterpreted dependencies. Compile errors and runtime errors also account for a substantial portion of failures, reflecting syntactic and logical flaws in the generated code. The other category may not always trigger an explicit error during execution but still fails to produce the expected results. Execution validation is essential for filtering out the synthetic low-quality code.}

\subsection{Canary Samples}
\label{sub:canary_samples}
As presented in Table~\ref{tab:canary_samples}, we construct five types of canary samples (e.g., Email, Name, IP Address, Password, Username). In these canary samples, the instructions are privacy-free task descriptions, while the corresponding code snippets contain the respective type of PII. We ensure that the PII instances in the canary samples do not appear in any other training samples. The constructed canary samples share similar task instructions and PII types with OSS-Instruct PII dataset. This design makes the test prompts more likely to trigger LLMs to generate code snippets containing the specific PII from the canary samples, more precisely simulating and testing the \toolname's risk of privacy leakage in reality.

\subsection{\rev{Robustness}}
\label{sub:robustness}
\rev{We conduct a robustness study of \toolname's training. We set random seeds in the fine-tuning steps, privacy-free code snippets synthesis step, and round-trip validation step of \toolname, along with training steps of baselines, to demonstrate the stable utility superiority under randomness. Table~\ref{tab:seed_variance} reports the pass@1 scores (mean $\pm$ standard deviation) of \toolname and the baselines under $\epsilon = 4$ when using Qwen2.5-Coder-7B as the base model. Even under different random seeds, \toolname consistently outperforms the other privacy-preserving baselines across all benchmarks, achieving improvements of up to 11.4\% on instruction-following benchmarks and 10.3\% on code-completion benchmarks.}

\subsection{Ablation Study}



\label{sub:ast vs noast}


Figure~\ref{fig:humaneval_examples} presents the examples of code generated in the HumanEval benchmark using the same prompt, where models are triggered to generate more complex code. Figure~\ref{fig:mbpp_examples} presents examples of relatively simple code generated in the MBPP benchmark. We use DS-Coder-6.7B~\cite{guo2024deepseekcoderlargelanguagemodel} as the evaluated LLM. While both \toolname and NonASTPrivCode pass the MBPP test case, only \toolname successfully passes the HumanEval test case. Examples show that \toolname has a stronger ability to generate more complex code, with higher structural and syntactic accuracy compared to the NonASTPrivCode.

\subsection{\rev{Analysis of Validation Filters}}
\label{sub:analysis_validation}
\rev{This section provides a detailed analysis of the acceptance rates of the execution and round-trip validation filters, the sensitivity analysis of the BERTScore threshold $\tau_s$ in the round-trip filter, and a taxonomy contrasting syntax-level and semantic-level failures. The relaxed setting in Section~\ref{sub:utility} generates 20 candidate descriptions per code snippet, whereas the stringent setting in Sections~\ref{sub:ablation}
and~\ref{sub:Hyper} generates only one, providing fewer opportunities to obtain a correctly formatted description. In both settings, we evaluate only the first candidate
matching the required output format and retain the code snippet only if its BERTScore meets the specified threshold; snippets with no format-matching candidate are discarded.}

\noindent\textbf{\rev{Acceptance Rates of Validation Filters.}} \rev{Under the experimental setup of Section~\ref{sub:utility}, we generate 55{,}500 synthetic samples. 
After applying the execution validation, 31{,}644 samples remain, with 57\% acceptance rate, indicating that a substantial portion of model outputs contain syntax errors, runtime failures, missing dependencies, or other execution-level issues, referring to Appendix~\ref{sub:execution_validation_appendix}. 
Subsequently, the round-trip validation further filters these 31{,}644 samples and retains 15{,}311, under the BERTScore threshold of 0.88, with 48\% acceptance rate, resulting in only 27.6\% of the total synthetic samples surviving both filters. 
This demonstrates that both stages of validation are necessary to ensure syntactic correctness and semantic fidelity.}

\noindent\textbf{\rev{Sensitivity to the BERTScore Threshold $\tau_s$.}}
\rev{We conduct hyper-parameter study against the round-trip validator under thresholds $\tau_s \in \{0.60, 0.75, 0.88, 0.95\}$ to understand its impact on semantic filtering strictness and final utility of \toolname. The acceptance rate of round-trip validation directly reflects the sensitivity to semantic filtering strictness. We observe that lower thresholds substantially increase acceptance but introduce semantic drift: $\tau_s=0.60$ admits 25{,}491 samples (46\% acceptance), and $\tau_s=0.75$ retains 19{,}842 samples (36\% acceptance), yet both settings allow many semantically inconsistent or loosely aligned outputs. In contrast, a very high threshold such as $\tau_s=0.95$ is overly strict and keeps only 5{,}985 samples (11\% acceptance), sharply reducing usable data. The mid-range threshold $\tau_s=0.88$, which retains 15{,}311 samples (28\% acceptance), achieves the best balance between preserving dataset scale and maintaining semantic fidelity.

Figure~\ref{fig:hyper_tau} shows that the final utility of \toolname initially increases with $\tau_s$ and then decreases beyond an optimal point. Under the studied thresholds $\{0.60, 0.75, 0.88, 0.95\}$, $\tau_s=0.88$ achieves the best trade-off. While a high threshold such as $\tau_s=0.95$ ensures semantic correctness of the filtered data, it reduces diversity and dataset size, yielding high pass@1 scores across four benchmark tests $\{49.4, 66.9, 43.0, 56.5\}$. Conversely, a low threshold such as $\tau_s=0.60$ retains amount low-quality semantically misaligned samples, resulting in substantial performance drops $\{32.9, 56.9, 40.3, 46.4\}$. We clarify that excessively high thresholds lead to maintain samples lacking diversity and then increase the risk of overfitting, reducing generalization performance. Therefore, selecting an appropriate value of $\tau_s$ is critical to balancing the semantic correctness and diversity of the dataset.}

\noindent\textbf{\rev{Filter Failure Taxonomy.}}
\rev{The execution validation primarily removes \textit{syntax-level} failures, including compile errors, runtime exceptions, missing dependencies, and language mismatches (as detailed in Appendix~\ref{sub:execution_validation_appendix}). 
In contrast, round-trip validation targets \textit{semantic-level} deviations rather than syntactic failures. It captures cases where the generated code diverges from the prompt in functionality, reproduces only part of the intended behavior, or is loosely related to the instruction, resulting in low round-trip similarity. Unlike execution validation, these failures reflect shortcomings in semantic fidelity rather than syntactic correctness.}

\subsection{Case Study of Private Information Protection}
\label{sub:pii case study}
Table~\ref{tab:pii_leakage_examples} presents a case study showing how \toolname can avoid leaking private information from the training dataset, while NonDPFT memorizes and reproduces PIIs in training code snippets verbatim or partially. When we carefully design a prompt that is similar in form or semantics to a sample prompt from the training dataset, referencing the training prompt in Table~\ref{tab:pii_leakage}, and which may contain fabricated identifiers resembling real PIIs. Instructed by this prompt, the NonDPFT model reproduces private information from the corresponding training code snippet in Table~\ref{tab:pii_leakage}, either verbatim or partially. Additionally, other private information that the model memorized in other training data samples are deemed similar, are also leaked by NonDPFT. In contrast, \toolname effectively avoids leaking privacy under $\epsilon=\{1,4,10\}$ while maintaining the correctness of the generated code.

\end{document}